\documentclass[epsfig,12pt]{article}
\usepackage{epsfig}
\usepackage{graphicx}
\usepackage{hyperref} 
\usepackage{multirow}
\graphicspath{{pics/}}
\DeclareGraphicsExtensions{.pdf,.png,.jpg}

\usepackage{csquotes}

\usepackage{array}
\usepackage{amsmath}
\usepackage{amssymb}

\newcommand{\beq}{\begin{equation}}   
	\newcommand{\eeq}{\end{equation}}
\newcommand{\beqn}{\begin{eqnarray}}   
	\newcommand{\eeqn}{\end{eqnarray}}

\usepackage{xcolor}

\begin{document}
	
	\hypersetup{%
		%colorlinks=false,% hyperlinks will be black
		linkbordercolor=blue,% hyperlink borders will be =color
		%pdfborderstyle={0 0 0.1}% 
	}
	
	\unitlength = 1mm
	
	\def\de{\partial}
	\def\Tr{ \hbox{\rm Tr}}
	\def\const{\hbox {\rm const.}}  
	\def\o{\over}
	\def\im{\hbox{\rm Im}}
	\def\re{\hbox{\rm Re}}
	\def\bra{\langle}\def\ket{\rangle}
	\def\Arg{\hbox {\rm Arg}}
	\def\Re{\hbox {\rm Re}}
	\def\Im{\hbox {\rm Im}}
	\def\diag{\hbox{\rm diag}}
	
	%%%%%%%%%%%%%%%%%%%%%%%%%%%%%%%%%%%%%%%%%%%%%%%%%%%%%%%%%%%%%%%%%%%%
	
	\def\QATOPD#1#2#3#4{{#3 \atopwithdelims#1#2 #4}}
	\def\stackunder#1#2{\mathrel{\mathop{#2}\limits_{#1}}}
	\def\stackreb#1#2{\mathrel{\mathop{#2}\limits_{#1}}}
	\def\Tr{{\rm Tr}}
	\def\res{{\rm res}}
	\def\Bf#1{\mbox{\boldmath $#1$}}
	\def\balpha{{\Bf\alpha}}
	\def\bbeta{{\Bf\beta}}
	\def\bgamma{{\Bf\gamma}}
	\def\bnu{{\Bf\nu}}
	\def\bmu{{\Bf\mu}}
	\def\bphi{{\Bf\phi}}
	\def\bPhi{{\Bf\Phi}}
	\def\bomega{{\Bf\omega}}
	\def\blambda{{\Bf\lambda}}
	\def\brho{{\Bf\rho}}
	\def\bsigma{{\bfit\sigma}}
	\def\bxi{{\Bf\xi}}
	\def\bbeta{{\Bf\eta}}
	\def\d{\partial}
	\def\der#1#2{\frac{\d{#1}}{\d{#2}}}
	\def\Im{{\rm Im}}
	\def\Re{{\rm Re}}
	\def\rank{{\rm rank}}
	\def\diag{{\rm diag}}
	\def\2{{1\over 2}}
	\def\ntwo{${\mathcal N}=2\;$}
	\def\nfour{${\mathcal N}=4\;$}
	\def\none{${\mathcal N}=1\;$}
	\def\ntwot{${\mathcal N}=(2,2)\;$}
	\def\ntwoo{${\mathcal N}=(0,2)\;$}
	\def\x{\stackrel{\otimes}{,}}

	\newcommand{\cpn}{CP$(N-1)\;$}
	\newcommand{\wcpn}{wCP$_{N,\widetilde{N}}(N_f-1)\;$}
	\newcommand{\wcpd}{wCP$_{\widetilde{N},N}(N_f-1)\;$}
	\newcommand{\wcpN}{$\mathbb{WCP}(N,N)\;$}
	\newcommand{\wcpK}{$\mathbb{WCP}(K,K)\;$}
	\newcommand{\wcpt}{$\mathbb{WCP}(2,2)\;$}
	\newcommand{\wcpf}{$\mathbb{WCP}(4,4)\;$}
	\newcommand{\wcpo}{$\mathbb{WCP}(1,1)\;$}
	\newcommand{\wcp}{$\mathbb{WCP}(N,\tilde N)\;$}
	\newcommand{\vp}{\varphi}
	\newcommand{\pt}{\partial}
	\newcommand{\tN}{\widetilde{N}}
	\newcommand{\ve}{\varepsilon}
	\renewcommand{\theequation}{\thesection.\arabic{equation}}
	
	\newcommand{\sun}{SU$(N)\;$}
	
	\setcounter{footnote}0
	%\begin{titlepage}
	%\renewcommand{\thefootnote}{\fnsymbol{footnote}}
	
	\vfill
	
	%%%%%%%%%%%%%%%%%%%%%%%%%%%%%%%%
	\begin{titlepage}

		\vspace*{-3cm}
				\begin{flushright}
						{\footnotesize 
								FTPI-MINN-25-09 \\ UMN-TH-4503/25
						}
				\end{flushright}

		\begin{center}
			{  \Large \bf  
				
				Critical Non-Abelian Vortex String
				\\[2mm]
				and 2D \ntwo Black Hole
			}
			
			% Critical Non-Abelian Vortex String and 2D N=2 Black Hole
			% E. Ievlev, A. Marshakov, G. Sumbatian, and A. Yung
			
			\vspace{5mm}
			%\vspace{1mm}
			
			{\large  \bf E.~Ievlev,$^{\,a}$ A.~Marshakov,$^{\,\,b,c}$ G.~Sumbatian,$^{\,d}$ and  A.~Yung$^{\,d,e}$}
		\end{center}

		% AFFILIATIONS
		\begin{center}
			
			{\it  $^{a}$William I. Fine Theoretical Physics Institute,
				University of Minnesota,
				Minneapolis, Minnesota 55455, USA}\\
			
			$^{b}${\it Dept. Math., HSE University,
				Moscow 119048, Russia
			}\\
			$^{c}${\it Theory Department of LPI,
				Moscow 119991, Russia}
			
			$^{d}${\it National Research Center ``Kurchatov Institute'',
				Petersburg Nuclear Physics Institute, Gatchina, St. Petersburg
				188300, Russia}\\
			
			{\it $^{e} $HSE University, St. Petersburg,
				194100, Russia}
			
		\end{center}

		%\vspace{5mm}
		
		\begin{center}
			{\large\bf Abstract}
		\end{center}
		\noindent
		It has been shown that the non-Abelian vortex string in 4D $\mathcal{N}=2$ supersymmetric QCD (SQCD) with the U(2) gauge group and $N_f=4$ flavors becomes a critical superstring. Its 10D target space is a product of the flat 4D space and an internal noncompact Calabi-Yau threefold, namely, the conifold. It was also shown that the Coulomb branch of the associated string sigma model, which opens up at strong coupling, can be described by $\mathcal{N}=2$ Liouville theory. We continue here the study of the recently proposed mass deformation of the U(2) theory with $N_f=4$, interpolating to SQCD with the U(4) gauge group and $N_f=8$ quarks, by analyzing the mass-deformed $\mathcal{N}=2$ Liouville theory on the string world sheet, and show that it is always described by the trumpet geometry of the target space, which is $T$-dual to the 2D $\mathcal{N}=2$ supersymmetric black hole. We use this correspondence to find the low-lying hadron spectrum in the deformed SQCD, and explain the expected increase in the number of hadronic states in the theory with more gauge fields and quarks by considering the near-Hagedorn behavior of the 2D black hole.
		
		%\vspace{2cm}
		
	\end{titlepage}
	
	\newpage
	
	\tableofcontents

	\newpage
	
	\section{Introduction}
	
	Non-Abelian vortices were first found in  four-dimensional 
	(4D) \ntwo  supersymmetric QCD (SQCD) 
	with the gauge group U$(N)$ and $N_f \ge N$ flavors of quarks
	\cite{HT1,ABEKY,SYmon,HT2}. The non-Abelian vortex string is 1/2
	BPS (Bogomol'nyi-Prasad-Sommerfield) saturated and therefore has \ntwot supersymmetry on its world sheet.
	In addition to four translational moduli,  the non-Abelian string carries orientational  moduli, as well as the size moduli if $N_f>N$
	\cite{HT1,ABEKY,SYmon,HT2} (see \cite{Trev,Jrev,SYrev,Trev2} for reviews). The dynamics of
	the internal orientational and size moduli of the non-Abelian  string
	are described by the effective two-dimensional (2D) sigma model on the string world sheet, the so-called \ntwot supersymmetric weighted $\mathbb{CP}$-model $\mathbb{WCP}(N,N_f-N)$.  
	
	It was shown in \cite{SYcstring}  that the non-Abelian solitonic  string in \ntwo SQCD  with the U(2) gauge group and $N_f = 2N=4$
	flavors of quark hypermultiplets behaves as a critical superstring~\footnote{ By critical superstring we mean canonical superstring with \none local supersymmetry in the world-sheet (super)gravity sector. The \ntwot supersymmetric world-sheet models, including supersymmetric \ntwo Liouville belong to the matter sector of string theory.}.
	For $N_f=2N$
	the world sheet sigma model $\mathbb{WCP}(N,N)$ becomes conformal, and additionally for $N=2$ the 
	six orientational/size moduli can be combined with 
	four translational moduli to form a ten-dimensional target space for a critical 
	superstring \cite{SYcstring,KSYconifold}. In this case the target space of the world sheet  theory is
	$\mathbb{R}^4\times Y_6$, where $Y_6$ is a noncompact six-dimensional Calabi-Yau (CY) manifold, the conifold \cite{Candel}, see \cite{NVafa} for a review. The theory of the critical vortex
	string was identified as the type IIA superstring theory \cite{KSYconifold}, and the
	spectrum of low-lying closed string excitations was found in  \cite{KSYconifold,SYlittles}.
	
	%A version of the string-gauge duality for 4D SQCD was proposed \cite{SYcstring}: at
	%weak coupling this theory is in the Higgs phase and can be described in terms
	%of quarks and Higgsed gauge bosons, while at strong coupling hadrons of this theory can be understood as closed string states formed by the non-Abelian vortex string. We  call this approach a ''solitonic string-gauge duality''.
	
	Most of the massless  and massive  string modes have  non-normalizable wave functions over the conifold $Y_6$, i.e. they are not localized in 4D 
	and cannot be interpreted as dynamical states in 4D theory, in particular, there are no massless 4D gravitons in the physical spectrum  \cite{KSYconifold}.
	However, an excitation associated with the deformation of the complex structure modulus $b$ of $Y_6$ has a (logarithmically) normalizable wave function   and was  interpreted  as a massless baryon in the spectrum of hadrons of 4D \ntwo SQCD  \cite{KSYconifold}.
	
	To analyze the massive states, a different approach was chosen,
	similar to that used for little string theories  (see  \cite{Kutasov} for a review). It is  based
	on the  equivalence \cite{GVafa} between the 
	critical string on the conifold and a noncritical $c=1$ string containing the Liouville 
	field and a compact scalar  at 
	the self-dual radius (united into a complex scalar of \ntwo Liouville theory \cite{Ivanov,KutSeib})~\footnote{In \cite{GVafa} this equivalence was shown for topological versions of the string theories.}.  
	Later a similar correspondence was proposed (and treated as a holographic AdS/CFT-type duality)
	for the critical string on certain  noncompact CY spaces with an isolated singularity in the so-called double scaling limit, and a noncritical $c=1$ string with an additional 
	Landau-Ginzburg
	\ntwo superconformal system  \cite{GivKut,GivKutP}. In the conifold case, this extra Ginzburg-Landau
	conformal field theory (CFT) is absent.
	The above equivalence was used in \cite{SYlittles,SYlittmult} to find the low-lying spectrum of hadrons in
	4D \ntwo SQCD with gauge group U(2) and $N_f=4$ quark flavors.
	
	Recently, this equivalence was demonstrated in a more direct way. Namely, it was shown in \cite{GIMMY}
	that  Coulomb branches of  world sheet $\mathbb{WCP}(N,N)$ models  on noncompact toric CY manifolds,
	%with an isolated singularity 
	which open up at strong coupling, can be
	described by \ntwo Liouville theory with $N$-dependent background charge. 
	% This was shown first in the large $N$ approximation and then extrapolated  to an exact equivalence.
	Using the description of the theory of the critical non-Abelian string in terms of \ntwo Liouville theory, the program of  interpolation between SQCDs with different gauge groups and numbers of quark flavors was initiated in \cite{Y_mass_Liouville}. The motivation is to broaden the class of 4D \ntwo  SQCDs where the hadron spectrum can be described using the string theory of the  non-Abelian string.

	The idea was that, to study this interpolation, one can  introduce quark masses in \ntwo SQCD and by changing  mass parameters decouple certain quark flavors.
	In particular,  one can interpolate between \ntwo SQCD with gauge group U(2) and $N_f=4$ (which supports a critical non-Abelian  string) and \ntwo SQCD with  gauge group U(4) and $N_f=8$ quark flavors ''integrating four quark flavors  in'' by reducing their mass parameters \cite{Y_mass_Liouville}.

	Of course, the quark masses break 
	the conformal invariance on the world sheet, and the mass-deformed theory cannot be used directly for the  string quantization. Instead, to find a true  string vacuum, the effective supergravity equations of motion can be solved with appropriate initial conditions
	% (at large values of the Liouville coordinate) 
	associated with the mass deformation \cite{Y_mass_Liouville}.
	
	In this paper, we continue these studies and show that actually the mass-deformed background found in \cite{Y_mass_Liouville} is $T$-dual to the  \ntwo supersymmetric 2D
	black hole with cigar geometry \cite{Wbh}, which is the  \ntwo SL($2, \mathbb{R}$)/U(1) coset Wess-Zumino-Novikov-Witten (WZNW) model \cite{GVafa,GivKut,MukVafa,OoguriVafa95}. In fact, the metric found in  \cite{Y_mass_Liouville}
	can be reduced to the metric of a trumpet, first discussed in \cite{Giveon} and used in \cite{DijVerVer} for study of winding modes in the 2D black hole geometry of the cigar.
	
	It is known that the original  \ntwo Liouville theory deformed with a Liouville superpotential has a mirror description in terms of a \ntwo 2D black hole \cite{HoriKapustin}. Using this, we argue that the combination of both superpotential  and mass deformations leads to the black hole with the same  cigar geometry, while the mass of the 2D black hole depends now on both deformation parameters. Namely, we conjecture that the total black hole mass is the sum of masses associated with each of the above  deformations. This conjecture is one of the key points of our paper, however actually we use it only in the limit  when the mass deformation dominates. %, which is  described by the $T$-duality.

	This allows us to find the low-lying hadron spectrum of the \ntwo SQCD with the U(4) gauge group and $N_f=8$ quark flavors. In fact,  it turns out that  the form of the spectrum of string states (which determines  
	the hadron spectrum in 4D SQCD) does not depend on the mass deformation, and only the number of states is changed. We also use the field theory arguments on the SQCD side to explain this surprising effect.  
	We explain the expected increase in the number of hadronic states in SQCD, which has  more quarks, by considering the near-Hagedorn behavior of the 2D black hole.
	
	The paper is organized as follows. In Sec.~\ref{sec:NAstring} we briefly review the 4D \ntwo
	supersymmetric SQCD and the world sheet theory of the non-Abelian string.
	In particular, we review results of \cite{GIMMY} which show that the Coulomb branch of the world sheet \wcpN model is described by \ntwo Liouville theory and describe the low-lying  4D spectrum in terms of string states in the world-sheet theory for $N=2$. In Sec.~\ref{sec:gravity} we consider the mass deformation and show that the mass-deformed solution, found in \cite{Y_mass_Liouville}, is $T$-dual to the 2D black hole with cigar geometry. We also 
	discuss the Liouville theory with both superpotential and mass deformations switched on, and formulate a conjecture how the deformation parameters are related to the background of the dual theory with cigar geometry. 
	%{\color{red} This conjecture is one of the key points of our paper, however the main physical results actually depend only on its obvious part.} 
	In Sec.~\ref{sec:hyper} we solve the Schr\"odinger equation for string excitations in the mass-deformed background, using the effective gravity approach, and find the discrete spectrum,
	the result confirms the statement about $T$-duality from Sec.~\ref{sec:gravity}. In Sec.~\ref{sec:SQCD_interpretation} we use the field theory arguments from SQCD to explain the surprising 
	behavior of the 4D spectrum, predicted by string theory.  In Sec.~\ref{sec:Hagedorn} we finally study the near-Hagedorn behavior of the 2D black hole to explain the expected increase in hadron states
	in the 4D SQCD with more quark flavors. Sec.~\ref{sec:Conclusions} contains our conclusions,
	while the Appendix~\ref{app:sing_cont} deals with an analytic continuation of the singular solution to the Schr\"odinger equation for string excitations.

	\section {Non-Abelian vortex string and $\mathcal{N}=2$ Liouville theory}
	\label{sec:NAstring}
	\setcounter{equation}{0}

	In this section we are going to review some basic facts about vortex strings in ${\mathcal N}=2\;$ SQCD.
	The goal is to make the reader familiar with the world-sheet theory and to explain how the hadron spectrum emerges in this construction, which will be generalized later in this work.

	\subsection{Four-dimensional \boldmath{${\mathcal N}=2\;$} 
		SQCD}
	\label{sec:SQCD}

	Our starting point is the $\mathcal{N}=2$ SQCD in 4D with eight supercharges and gauge group U(1)$\times$SU$(N)$; see, for example, \cite{SYrev} for a detailed review of this theory.
	We take the number of fundamental quark hypermultiplets to be $N_f = 2 N$; with this choice, the $\beta$-function of the 4D SQCD is zero and the 4D coupling does not run.  However, the conformal
	invariance of the 4D theory is explicitly broken by 
	the Fayet-Iliopoulos (FI)  term  \cite{FI} with FI parameter $\xi$, which defines the vacuum expectation values (VEVs) of quarks. The FI
	parameter is not renormalized.
	
	At weak coupling $g^2\ll 1$, this theory is in the Higgs phase. 
	In a vacuum where the first $N$ quark flavors are massless at zero $\xi$, the matrix of adjoint scalars of the $\mathcal{N}=2$ vector multiplet develops VEV of the form
	\begin{equation}
		\langle a \rangle = -
		%	\frac{1}{\sqrt{2}}
		\text{diag} (m_1 \,, \ldots \,, m_N) \,,
		\label{avev}
	\end{equation}
	where  $m_A$ ($A=1,..., N_f$) are bare quark masses. Adjoint condensates \eqref{avev} break U$(N)$ gauge group down to U(1)$^N$, with the masses of the off-diagonal gauge bosons given by 
	$|m_k-m_l|$ ($k,l=1,..., N$), while the quark masses of $q^{kA}$ and $\tilde{q}_{Ak}$ (two complex scalars of the $\mathcal{N}=2$ hypermultiplet) are equal to $|m_k-m_A|$. 
	
	At nonzero $\xi$, first $N$ squarks also develop VEVs,
	\begin{equation}
		\langle q^{kA}\rangle =\sqrt{
			\xi}\, \delta^{kA}, \qquad \langle \tilde{q}_{Ak} \rangle =0\qquad k=1,...,N ,\; A=1,...,N_f.
		\label{qvev}
	\end{equation}
	%
	%Adjoint VEVs break U$(N)$ gauge group down to U(1)$^N$ which is further Higgsed by squarks.
	%Masses of off-diagonal gauge bosons are given by 
	%$M^k_l \sim |m_k-m_l|$. 
	%If certain quark masses coincide with each other, adjoint VEVs leave  certain non-Abelian subgroups of U$(N)$ unbroken, see below. 
	%Masses of $q^{kA}$ and $\tilde{q}_{Ak}$ quarks are equal to $|m_k-m_A|$ (they vanish for the squark components with $A = k \leqslant N$).
	%
	%
	These quarks' VEVs break the U$(N)$ gauge group, Higgsing  all gauge bosons. The Higgsed gauge bosons combine with the screened quarks to form long \ntwo multiplets with mass $\sim g\sqrt{\xi}$ in the limit of zero quark masses.

	In this limit, the global flavor SU$(N_f)$ is also broken down by quark VEVs to the so-called color-flavor
	locked group. The resulting global symmetry is
	\beq
	{\rm SU}(N)_{C+F}\times {\rm SU}(N)\times {\rm U}(1)_B,
	\label{c+f}
	\eeq
	see \cite{SYrev} for more details. 
	The unbroken global U(1)$_B$ factor above is identified with a baryonic symmetry. 
	%Note that 
	%what is usually identified as the baryonic U(1) charge is a part of  our 4D theory  gauge group.
	% ``Our" U(1)$_B$
	%is  an unbroken by squark VEVs combination of two U(1) symmetries;  the first is a subgroup of the flavor 
	%SU$(N_f)$, and the second is the global U(1) subgroup of U$(N)$ gauge symmetry.
	
	In the Higgs phase, quarks are screened, while monopoles are confined by non-Abelian strings. 
	In fact, in  U$(N)$ theories, confined monopoles are junctions of two distinct elementary non-Abelian strings, see \cite{SYrev} for a review. In particular, baryons represent closed \textquote{necklace}
	configurations of monopoles on the string. At weak coupling, all these stringy hadrons are heavy and decay into perturbative states; however, at strong coupling, the theory enters the so-called instead-of-confinement phase \cite{SYi_of_c,ISY_b_baryon}. In this phase quarks and gluons decay into monopole-antimonopole pairs and we are left with hadrons formed by monopoles confined by non-Abelian strings.
	
	Below we  assume that $N$ is even, $N=2K$, where $K$ is integer
	and  consider a special choice of quark masses,
	\beq
	\tilde{m}_A =m_A, \qquad A=1,...,N,
	\label{extra_masses}
	\eeq
	where $\tilde{m}_A\equiv m_{A+N}$, $A=1,...,N$.  This  ensures that \textquote{extra} quarks  with $A=(N+1),...,2N$ have the same masses as the  first $N$ ones. Moreover,  we assume that quark masses  are taken to be
	\begin{equation}
		\{ m_A \}_{A=1}^{N_f} = \{ \underbrace{0\,,\ldots\,,0}_{N/2} \,,  \underbrace{M\,,\ldots\,,M}_{N/2} \,,  \underbrace{0\,,\ldots\,,0}_{N/2} \,,  \underbrace{M\,,\ldots\,,M}_{N/2}   \}
		\label{mass_split}
	\end{equation}
	with $M$ being the deformation parameter.
	In the large $M$ limit, the original SQCD splits into two sectors with mutual interactions suppressed by the scale $1/M$,
	\begin{equation}
		[SU(2K) + (N_f=4K)] \xrightarrow{ M \to \infty } [SU(K) + (N_f=2K)] \times [SU(K) + (N_f=2K)]
		\label{M_to_infty}
	\end{equation}
	(see \cite{Y_mass_Liouville} for details).
	In the opposite limit $M \to 0$, we recover the full SU$(2K)$ with $N_f=4K$ fundamental flavors.
	Thus, varying the parameter $M$, we can interpolate between these two theories.
	
	The case with $K=2$, when both SQCDs in the rhs of \eqref{M_to_infty} support critical non-Abelian strings, will be the main subject of this study.

	\subsection{World-sheet theory of the non-Abelian  string}
	\label{sec:wcp}
	
	\subsubsection{\wcpN model}
	
	The presence of the color-flavor locked group SU$(N)_{C+F}$ in 4D \ntwo SQCD with gauge group U$(N)$ is the reason for the formation of 
	non-Abelian vortex strings \cite{HT1,ABEKY,SYmon,HT2}.
	The most important feature of these vortices is the presence of the  orientational zero modes described by complex fields $n^i$, $i=1,...,N$ living on the 2D world sheet.
	In our case, the number of quark flavors exceeds the number of colors; the solitonic vortices become semilocal and acquire extra size moduli \cite{AchVas},
	described by $N_f - N = N$ complex fields $\rho^j$ with $j=1,...,N$; see \cite{HT1,HT2,AchVas,SYsem,Jsem,SVY}. 
	In \ntwo SQCD the flux-tube strings are 1/2 BPS saturated and preserve \ntwot supersymmetry with four supercharges on the world sheet.
	Their tension is determined exactly by the FI parameter,
	\beq
	\tau=2\pi \xi
	\label{ten}
	\eeq 
	The effective theory on the string world sheet is 2D \ntwot supersymmetric \wcpN model, defined  as a low-energy limit of the  U(1) gauge theory \cite{W93},
	%\footnote{``W'' here stands for weighted.} 
	with twisted-mass deformation
	\begin{equation}
		\begin{aligned}
			&S = \int d^2 x \left\{
			\left|\nabla_{\alpha} n^{i}\right|^2 
			+\left|\widetilde{\nabla}_{\alpha} \rho^j\right|^2
			-\frac1{4e_0^2}F^2_{\alpha\beta} + \frac1{e_0^2}\,
			\left|\pt_{\alpha}\sigma\right|^2 + \frac1{2e_0^2}\,D^2- \frac{\Theta}{2\pi}F_{01}
			\right.
			\\[3mm]
			&-
			\left.\left|\sqrt{2}\sigma +m_i\right|^2 \left|n^{i}\right|^2 - \left|\sqrt{2}\sigma +\tilde{m}_j\right|^2 \left|\rho^j\right|^2
			+D\left(\left|n^{i}\right|^2-\left|\rho^j\right|^2 - {\rm Re}\,\beta \right)\right\},
			\\[3mm]
			&  \alpha,\beta=1,...,2\,,\quad i,j=1,...,N, 
		\end{aligned}
		\label{wcpNN}
	\end{equation}
	see review \cite{SYrev} for details.  The fields $n^{i}$ and $\rho^j$ have
	charges  $+1$ and $-1$, respectively 
	\begin{equation}
		\nabla_{\alpha}=\pt_{\alpha}-iA_{\alpha}\,,
		\qquad 
		\widetilde{\nabla}_{\alpha}=\pt_{\alpha}+iA_{\alpha}\,.	
		\label{cov_derivatives}
	\end{equation}
	Twisted masses of the $n^i$ and $\rho^j$ fields coincide with the masses $m_i$ and $\tilde{m_j}$ of the  4D quarks. 
	The complex scalar $\sigma$ is a superpartner of the U(1) gauge field $A_{\alpha}$ and $D$ is the auxiliary field in the vector
	supermultiplet. These fields can be written in terms of  the twisted chiral superfield $\Sigma$ \cite{W93}~\footnote{Here spinor indices are written as subscripts, say 
		$\theta^L=\theta_R$, $\theta^R= -\theta_L$. We also define the twisted measure 
		$d^2 \tilde{\theta} = \frac12\,d \bar{\theta}_{L} d\theta_{R}$ to ensure that 
		$\int d^2 \tilde{\theta}\, \tilde{\theta}^2 = \int d \bar{\theta}_{L} d\theta_{R}\, \theta_R \bar{\theta}_L=1$.} 
	\beq
	\Sigma = \sigma +\sqrt{2}\theta_R \bar{\lambda}_L -\sqrt{2}\bar{\theta}_L \lambda_R
	+\sqrt{2}\theta_R\bar{\theta}_L (D-iF_{01}).
	\label{Sigma}
	\eeq
	The complexified inverse coupling in \eqref{wcpNN},   
	\begin{equation}
		\beta = {\rm Re}\,\beta + i \, \frac{\Theta}{2 \pi} \,,
		\label{beta_complexified}	
	\end{equation}
	is defined via the  2D FI term  (twisted superpotential)
	\beq
	-\frac{\beta}{2}\,\int d^2 \tilde{\theta}\sqrt{2}\,\Sigma = - \frac{\beta}{2}\,(D-iF_{01})
	\label{Sigma_sup}.
	\eeq

	For our purposes, the following facts about this 2D theory will be most essential:
	\begin{enumerate}
		
		\item 
		The beta function for this sigma model vanishes, and this 2D theory is conformal in the zero-mass limit.
		Therefore, its target space is Ricci-flat and [being K\"ahler due to \ntwot supersymmetry] represents  a (noncompact) Calabi-Yau manifold.
		%The corresponding central charge (in terms of the bulk number of colors $N$) is given by
		%%\beq
		%%\hat{c}_{CY} \equiv \frac{c_{CY}}{3} = {\rm dim}_\mathbb{C}{\cal H} (4N-1-1)/2 = 2N-1 
		%%\label{cCY}
		%%\eeq
		%\beq
		%\hat{c}_{CY} = 2N-1 
		%\label{cCY}
		%\eeq
		As usual for such compactifications the \ntwot superconformal symmetry of 2D world-sheet theory ensures \ntwo spacetime supersymmetry in 4D after the Gliozzi–Scherk–Olive (GSO) projection.
		
		\item 
		The global symmetry group of the \wcpN sigma model coincides with the unbroken global group of the 4D SQCD \eqref{c+f}; see \cite{KSYconifold}.
		The fields $n^i$ and
		$\rho^j$ transform in   the representations 
		\beq
		\left(\,{\bf N}, \,{\bf 1}, \,\frac12 \right) \qquad \left(\,{\bf 1},\,{\bf N} , \,\frac12 \right)
		\label{n_rho_representations}
		\eeq
		respectively.
		Note that another U(1) symmetry, which rotates the $n$ and $\rho$ fields with opposite charges, is gauged.
		
		\item 
		2D-4D correspondence: the BPS spectrum of states in this 2D world sheet theory coincides with the BPS spectrum of 4D states in the quark vacuum \eqref{qvev} given by the exact Seiberg-Witten solution \cite{SW2} at $\xi=0$. This coincidence was observed in \cite{Dorey,DoHoTo} and explained
		later in \cite{SYmon,HT2}  using the picture of  4D monopoles confined by the non-Abelian string that are seen as kinks in the 2D world-sheet theory.
		
	\end{enumerate}
	
	The Calabi-Yau space associated with the massless \wcpt model is the conifold. It is six-dimensional, and, combined with the flat Minkowski 4D space,  forms a ten-dimensional  space required for a superstring to be critical
	\cite{SYcstring,KSYconifold}. In fact, we are interested in considering the \wcpt model at strong coupling $\beta=0$, the so-called ``thin string conjecture'' put forward in \cite{SYcstring,KSYconifold} implies that only at  vanishing $\beta$  \eqref{beta_complexified} we expect that the world sheet \wcpt model defines the consistent string theory for the solitonic  non-Abelian vortex in \ntwo 4D SQCD.
	
	At $\beta=0$, the conifold singularity ${\rm det}\, w^{ij} =0$, where $w^{ij} =n^i \rho^j$ is the set of U(1) gauge-invariant ''mesonic'' variables, can be alternatively smoothed by the deformation of the complex structure. Following \cite{NVafa}, this deformation can be described as 
	\beq
	{\rm det}\, w^{ij} =b,
	\label{deformedconi}
	\eeq
	where $b$ is a complex parameter. 
	Promoted to a 4D field, the deformation constant $b$ was interpreted as a scalar component of a  massless baryonic hypermultiplet  of 4D \ntwo QCD in  \cite{KSYconifold}. Its quantum numbers with respect to the global symmetry group \eqref{c+f} for  $N=2$ are
	determined by \eqref{deformedconi},
	\beq
	\left(\,{\bf 1}, \,{\bf 1}, \, 2 \right),
	\label{b_representation}
	\eeq 
	it is a singlet with respect to both SU(2) groups with $B(b)=2$ baryonic  charge \cite{KSYconifold}.
	
	The massless field $b$  can form a condensate. Thus, 
	we have a new Higgs branch in 4D \ntwo SQCD which develops only at the critical value of 
	the 4D coupling constant $\tau_{SW}=1$ associated with $\beta=0$ \cite{IYcorrelators}.

	\subsubsection{\ntwo Liouville theory}
	
	As we mentioned in the Introduction, it was recently shown in \cite{GIMMY} that the Coulomb branch of the 2D \wcpN world sheet model which opens up at strong coupling $\beta=0$ can be described by \ntwo Liouville theory (see \cite{Nakayama} for a review of Liouville theory in general).
	Its bosonic action reads
	\beq
	S_{\rm eff}=\frac{1}{4\pi}\int d^2 x\sqrt{h}
	\;\left(\frac1{2}\,h^{\alpha\beta}(\pt_{\alpha}\phi\pt_{\beta}\phi  +\pt_{\alpha}Y\pt_{\beta}Y)
	-\frac{Q}{2}\phi\, R^{(2)} \right),
	\label{Liouville}
	\eeq
	for the real Liouville scalar field $\phi$, supplemented by the real compact scalar $Y \sim Y+2\pi$. Here $R^{(2)}$ is the Ricci scalar for the world sheet metric
	$h_{\alpha\beta}$,  and  $h={\rm det} (h_{\alpha\beta})$.  The linear  dilaton in \eqref{Liouville},
	\beq
	\Phi =-\frac{Q}2\,\phi
	\label{linear_dilaton}
	\eeq
	contains the background charge $Q$ for the  Liouville field $\phi$, given by
	\beq
	Q(N)=\sqrt{2(N-1)}.
	\label{Q}
	\eeq
	The action in \eqref{Liouville} leads to the following holomorphic stress tensor of the bosonic part of the theory 
	\beq
	T= -\frac12\,\left[(\pt_z \phi)^2 + Q\, \pt_z^2 \phi + (\pt_z Y)^2\right], 
	\label{T--}
	\eeq
	so that the  central charge is \(c_L=1+3Q^2+2=3(1+Q^2)\).  For the $N=2$ conifold case, \eqref{Q} gives $Q=\sqrt{2}$,
	so the central charge of the internal part of the world sheet theory equals nine as required for criticality.
	
	The \ntwo Liouville interaction superpotential (see \cite{Nakayama}) 
	for the $N=2$ case has the form 
	\beq
	L_{int}= b \,\int d^2 \tilde{\theta}\,e^{-\frac{\phi +iY}{Q}}
	\label{Liouville_sup},
	\eeq
	where we promote the scalars $\phi$ and $Y$ to (twisted) chiral superfields, and $b$ is the conifold complex structure deformation parameter \eqref{deformedconi}. It comes from  the 2D FI term \eqref{Sigma_sup} in the \wcpt model when we use the relation
	\beq
	\sigma=\gamma\, e^{-\frac{\phi + iY}{Q}},
	\label{sigma}
	\eeq
	between the complex scalar $\sigma$ from the U(1) gauge multiplet of the $\mathbb{WCP}(2,2)$ model and the fields $\phi$ and $Y$ in the Liouville theory, where $\gamma = -\sqrt{2}\,b/\beta$. Note that we must take the constant $\gamma$ singular in the limit $\beta\to 0$ in order to keep the Liouville superpotential \eqref{Liouville_sup} finite, see \cite{GIMMY}
	for details.
	%~\footnote{The fact that the Liouville interaction is given by a twisted superpotential  is just a matter of conventions since there are no untwisted chiral fields in %the effective theory.}.
	This superpotential is a marginal deformation of the \ntwo Liouville theory \eqref{Liouville}.

	To conclude this subsection, we note that  the dilaton has a linear dependence on the Liouville coordinate $\phi$, see \eqref{linear_dilaton}. Therefore, the string coupling constant $g_s=e^{\Phi}$ would become large at large negative $\phi$. On the other hand,  at nonzero $b$, the Liouville wall 
	prevents the field $\phi$ from penetrating into the region of large negative values. In fact, the maximum value of the string coupling is 
	$g_s\sim 1/|b|$ for $Q=\sqrt{2}$. In this paper, we keep $b$
	large to ensure that the string coupling is small and the string
	perturbation theory is reliable, see \cite{GivKut,SYlittmult}. In particular, one can use
	the tree-level approximation to obtain the string spectrum.
	In terms of 4D SQCD, keeping $b$ large
	means moving along the Higgs branch far away from the origin.

	\subsection{Primary operators on the world sheet }
	\label{sec:vertices}
	
	In this subsection, we review primary operators in the \ntwo Liouville theory. 
	For the $N=2$ [$Q =\sqrt{2(N-1)}= \sqrt{2}$] case, they describe physical string states interpreted as  hadrons in 4D SQCD; see \cite{SYlittles} for details.
	
	The primary operators at large $\phi$ have the form
	\cite{IYcorrelators,Teschner:1999ug,LSZinLST}
	\begin{equation}
		T_{j, m_L, m_R}
		\simeq
		%	\frac{1}{2j+1}
		e^{i Q(m_L Y_L - m_R  Y_R)}\left[e^{Q j \phi}+ R(j, m_{L,R} ; k) e^{-Q(j+1) \phi}\right]
		\label{genericj}
	\end{equation}
	where $Y_{L,R}$ correspond to the holomorphic and antiholomorphic parts of the compact scalar, with quantum numbers $m_{L,R}$,
	\beq
	m_L= \frac12(n_1+kn_2), \qquad  m_R= \frac12(n_1-kn_2), \qquad k=\frac{2}{Q^2},
	\label{m}
	\eeq
	related to integer winding ($n_1$) and  momentum ($n_2$) numbers, respectively. These operators should also obey $m_R=\pm m_L$ to have equal left and right conformal dimensions,
	\beq
	\Delta_{j,m} = \frac{Q^2}{2}\left\{m^2 - j(j+1)\right\} = \frac1{k}\left\{m^2 - j(j+1)\right\}\, ,
	\label{dimV}
	\eeq
	and unitarity requires \(\Delta_{j,m}> 0\).  For our 4D string applications, we set in \eqref{genericj} $m_L=-m_R\equiv m$.\footnote{This is the condition for the type IIA string, while for the type IIB string $m_{R}= m_{L}$ \cite{SYlittmult}. \label{foot:mLR}} This corresponds to the local or momentum operator with $n_2\neq 0$, $n_1=0$ (depending on \(Y=Y_L+Y_R\)) in the Liouville theory, and to the dual vortex or winding state in the mirror cigar picture. 
	
	The so-called reflection coefficient in \eqref{genericj}, given by \cite{Teschner:1999ug,LSZinLST,FatB}
	\begin{multline}
		R(j, m_L, m_R ; k) \\
		=
		\left[ \frac{1}{\pi} \frac{\Gamma\left(1+\frac{1}{k}\right)}{\Gamma\left(1-\frac{1}{k}\right)} \right]^{2j+1} 
		\frac{\Gamma\left(1-\frac{2 j+1}{k}\right) \Gamma(m_L+j+1) \Gamma(m_R+j+1) \Gamma(-2 j-1)}%
		{\Gamma\left(1+\frac{2 j+1}{k}\right) \Gamma(m_L-j) \Gamma(m_R-j) \Gamma(2 j+1)}
		\label{refl_GK}
	\end{multline}
	vanishes for values of $j$ and $m$  from the
	discrete spectrum
	\begin{equation}
		j=-\frac12, -1, -\frac32,..., \qquad m=\pm\{j, j-1,j-2,...\}.
		\label{discrete}
	\end{equation}
	for $k=1$ ($Q=\sqrt{2}$), which kills the rising exponential in \eqref{genericj},
	so that the primary operator gives a normalizable wave function at $j\le -1/2$; see \eqref{normalizable} below~\footnote{Another option is with ${\rm Re}\,j =-1/2$, when both exponentials are present in \eqref{genericj}, but they have the same normalization properties. This corresponds to the principal continuous representation with \(j=-\frac12+i\mathbb{R}\).}. 
	
	The spectrum \eqref{discrete}  was exactly determined using the
	mirror description  \cite{HoriKapustin} of the theory as a \ntwo ${\rm SL}(2,R)/{\rm U}(1)$ coset with a cigar geometry with the level
	\beq
	k=\frac{2}{Q^2}
	\label{k_Q}
	\eeq
	of the supersymmetric version of the Ka\v{c}-Moody algebra
	in \cite{MukVafa,DixonPeskinLy,Petrop,Hwang,EGPerry}, 
	see \cite{EGPerry-rev} for a review.
	
	To exclude the states with negative norm, one has to impose an extra restriction
	\cite{DixonPeskinLy,Petrop,Hwang,EGPerry,EGPerry-rev}, 
	\beq
	-\frac{k+1}{2}\leq j <0\,.
	\label{no_ghosts}
	\eeq
	These conditions are also seen from the reflection coefficient \eqref{refl_GK}, namely, they are associated with $\Gamma\left(1+\frac{2 j+1}{k}\right)$ and $\Gamma(-2 j-1)$,
	see \cite{IYcorrelators} for more details.
	
	Moreover, we look for string states with normalizable wave functions, 
	\beq
	\Psi_{j;m_L,m_R}(\phi,Y) = e^{-\Phi}T_{j, m_L, m_R}\ \stackreb{\phi\to\infty}{\sim} e^{Q(j+\frac{1}{2})\phi + i Q (m_L Y_L - m_R Y_R)}
	\label{psifun}
	\eeq
	to be interpreted as hadrons in 4D \ntwo SQCD. This requires
	\beq
	j\le -\frac12\,,
	\label{normalizable}
	\eeq
	where the borderline case $j=-\frac12$ is also included. 
	Thus, for our value $k=1$, we are left with only two options $j=-1/2$ and $j=-1$.\footnote{Note that $j=-1$ is included for $k=1$ as indicated in \eqref{no_ghosts} because 
		the pole of $\Gamma\left(1+\frac{2 j+1}{k}\right)$ is canceled by the  pole of 
		$\Gamma\left(1-\frac{1}{k}\right)$ from the prefactor.}

	\subsection{4D SQCD low-lying hadron spectrum}
	\label{sec: 4D spectrum}
	
	Dressing the tachyon vertex \eqref{genericj} with the 4D plane wave
	\beq
	\mathcal{T}_{j,m}=e^{ip_{\mu}x^{\mu}}\, T_{j,m,-m},
	\label{dressed_tachyon}
	\eeq
	we impose  the mass-shell condition 
	\beq
	\frac{p_{\mu}p^{\mu}}{2} + \Delta_{j,m} = \frac{1}{2}.
	\label{tachphys}
	\eeq
	Here we use dimensionless notation, imposing normalization $4\pi\tau=1$ ($\alpha' =2$, $\alpha' \equiv 1/(2\pi \tau$) and  Minkowski 4D metric ${\rm diag}(-1,1,1,1)$.
	%Note that the unity in the r.h.s. is shifted to 1/2 in the supersymmetric case due to the ghost contribution \cite{GivKut,SYlittles}. 
	The GSO projection restricts $2m$ for the operator \eqref{dressed_tachyon} at $k=1$ to be odd
	\cite{KutSeib,GivKut}, i.e.,
	\beq
	m=\frac12 +\mathbb{Z}.
	\label{oddn}
	\eeq
	Then the only possibility [see \eqref{discrete} and \eqref{no_ghosts}] is $j=-\frac12$.
	This determines the masses of the 4D scalars~ \footnote{These states are, of course, not tachyonic in 4D, but we will
		use the standard terminology and refer to them as “tachyons”.} to be \cite{SYlittles}
	\beq
	\frac{M_T^2}{2}=-\frac{p_{\mu}p^{\mu}}{2} = m^2 -\frac14 \,,\ \ \ \ m=\pm \frac12,\pm\frac32,\ldots
	\label{tachyonmass}
	\eeq
	In particular, the operator  \eqref{genericj} with $j=-1/2$ and $m_L=\pm 1/2$  has conformal dimension \eqref{dimV} 
	\beq
	\Delta_{j=-\frac12,m=\pm \frac12 } =\frac12 
	\label{Delta_b}
	\eeq
	so it is marginal and describes a massless string state in 4D. As was noted in \cite{SYlittles} this massless state corresponds to the complex structure modulus $b$ for the string compactification on the conifold. Two possible values of $m=\pm 1/2$ correspond to two real degrees of freedom of the complex scalar field $b$.
	The associated string state has a logarithmically normalizable wave function  in terms of the conifold radial coordinate  \cite{KSYconifold,Strominger_95}. On the Liouville side, this corresponds to the borderline normalization of the massless state \eqref{genericj} with $j=-\frac12$, $m=\pm \frac12$, see \cite{SYlittles} for details.

	Consider  now the 4D spin-2 states, corresponding to the vertex operators
	\beq
	\psi^{\mu}_L\psi^{\nu}_R\, 
	e^{ip_{\mu}x^{\mu}}\, T_{j;m,-m}\, ,
	\label{graviton}
	\eeq
	where $\psi^{\mu}_{L,R}$ are the world-sheet superpartners of the 4D coordinates $x^{\mu}$. 
	The GSO projection selects $2m$ to be even, $|m|=0, 1,2,...$ \cite{GivKut},
	thus we are left with the only value $j=-1$ in \eqref{discrete} and \eqref{no_ghosts}. 
	This leads to the following masses of spin-2 states \cite{SYlittles}:
	\beq
	M_V^2 = 2m^2, \qquad |m|=1,2,... .
	\label{gravitonmass}
	\eeq
	We  call 4D  states with  masses \eqref{gravitonmass} massive \textquote{gravitons},\footnote{Physical states in 4D SQCD come in \ntwo supermultiplets. This name underlines that the highest spin components of the corresponding supermultiplets have spin two, see \cite{SYlittmult}.} note that
	$m=0$, associated with the massless  4D graviton, is excluded.
	The momentum $m$ in the compact dimension is related to the baryonic charge as
	\cite{SYlittles,SYlittmult}
	\beq
	Q_B = 4m,
	\label{m-baryon}
	\eeq
	and all closed string states are baryons.

	\section {Mass deformation}
	\label{sec:gravity}
	\setcounter{equation}{0}

	In this section, we introduce  the mass deformation and  first review the solutions of  effective gravity equations for the mass-deformed superstring background found in \cite{Y_mass_Liouville}. Then,  we show that the mass deformed \ntwo Liouville theory is $T$-dual to the 2D supersymmetric  black hole with cigar geometry and reduce  its target space metric  to the metric of the trumpet. Finally, we discuss
	the theory with both the mass deformation and the Liouville superpotential  switched on.

	\subsection{Metric after  mass deformation}
	
	%In this section we implement 
	
	The bosonic part of the action of the type-II supergravity
	in string frame for the metric and dilaton is given by
	\beq
	S= \frac1{2\kappa^2}\int d^D x \, \sqrt{-G}\,e^{-2\Phi}\,\left\{ R + 4G^{MN}\pt_M\Phi\pt_N\Phi_N 
	%+{\rm const} 
	+ \cdots \right\},
	\label{gravity_action}
	\eeq
	where $G_{MN}$ is the $D$-dimensional metric, $M,N=1,...,D$.
	Here $2\kappa^2= (2\pi)^{(\frac{D}{2} -2)}g^2_s/\tau^{\frac{D}{2}-1}$.
	
	Einstein's equations  following from  \eqref{gravity_action} have the form
	\beq
	R_{MN}+ 2D_M D_N \Phi =0,
	\label{Einstein}
	\eeq
	while the equation for the dilaton reads
	\beq
	4G^{MN}\pt_M\Phi\pt_N\Phi -2G^{MN}D_M D_N \Phi +p =0,
	\label{dilaton_eq}
	\eeq
	where $p=\frac{D-10}{2}$ (in dimensionless units) is included when $D\neq 10$.	
	
	We assume that our space-time is a direct product of the flat 4D Minkowski space and an internal space which has a nontrivial metric,  
	\beq
	ds^2_{{\rm int}} = g(\phi) \left(d\phi^2 + dY^2\right) 
	\label{int_metric}
	\eeq
	associated with  mass-deformed \ntwo Liouville theory. Thus, $D=6$ and the solution, found in \cite{Y_mass_Liouville}, for the metric warp factor is
	\beq
	g(\phi)= \frac1{1-\frac{|b|^2}{|\mathcal{M}|^2}\,e^{- Q \phi}}= \frac1{1-e^{- Q (\phi-\phi_0)}},
	\qquad \phi_0 = -\frac1{Q}\log\frac{|\mathcal{M}|^2}{|b|^2}
	\label{g}
	\eeq
	and for the dilaton, 
	\beq
	\Phi (\phi)= -\frac{Q}{2} \phi +\frac12 \log g(\phi)=-\frac{Q}{2} \phi -\frac12\, 
	\log{\left[1-e^{- Q (\phi-\phi_0)}\right]}.
	\label{Phi_sol}
	\eeq
	We see that the warp factor of the metric and the dilaton are functions of the Liouville field $\phi$ and do not depend on $Y$.
	
	The solution \eqref{g} satisfies the initial condition
	\beq
	g(\phi)\approx 1+ \frac{|b|^2}{|\mathcal{M}|^2}\,e^{-\frac{2\phi}{Q}} +\cdots,
	\label{warp_ic}
	\eeq
	imposed  by the mass deformation only in our case $Q=\sqrt{2}$; see \cite{Y_mass_Liouville} for details. Here $\mathcal{M}= -\beta M/2$ is the rescaled parameter of the mass deformation, which we keep finite in the limit $\beta\to 0$.\footnote{This rescaling is due to the presence of the coefficient 
		$\gamma$ in the relation \eqref{sigma} missed in \cite{Y_mass_Liouville}.}
	
	Notice that in the limit $\mathcal{M}\to \infty$ solution \eqref{g}, \eqref{Phi_sol} turns into solution of \eqref{Einstein}, \eqref{dilaton_eq} with
	flat metric with $g(\phi)=1$ and the linear dilaton \eqref{linear_dilaton}. Einstein's equations are then valid trivially, while Eq. \eqref{dilaton_eq} is satisfied only when $Q=\sqrt{2}$ is matched with $p=-2$.

	Note  also that the first nontrivial term in the expansion of the warp factor \eqref{g} at large $\phi$ gives rise to the following
	deformation: 
	\beq
	(\pt_{z}\phi - i\pt_{z}Y) (\pt_{\bar{z}}\phi + i\pt_{\bar{z}}Y)  \,e^{- Q \phi}.
	\label{non-chiral_deform}
	\eeq
	of the free action. 
	This operator \eqref{non-chiral_deform} has $j=-1$, $m=0$, and  is marginal with conformal dimension $\Delta =(1,1)$. It is  the bosonic part of  so-called nonchiral marginal deformation of  \ntwo Liouville theory; see \cite{Nakayama} for a review. We see that \eqref{g}  and \eqref{Phi_sol} represent an exact solution for the mass  deformation, which is infinitesimally associated  with the nonchiral marginal operator \eqref{non-chiral_deform}.
	
	Putting together all terms, we get for the bosonic part of the world sheet action of the mass-deformed Liouville theory \cite{Y_mass_Liouville}
	\begin{equation}
		S_{\rm ws} =\frac{1}{4\pi}\int d^2 x \sqrt{h}\; \left\{\frac1{2}g(\phi)\left[ (\pt_{\alpha}\phi)^2 + (\pt_{\alpha}Y)^2\right] 
		+ \Phi(\phi)R^{(2)} + L_{int}\right\},
		\label{deformed_Liouville}
	\end{equation}
	where the metric warp factor $g(\phi)$ and the dilaton $\Phi(\phi)$ are given by \eqref{g}  and \eqref{Phi_sol}, while the Liouville
	superpotential  still takes the form \eqref{Liouville_sup}
	since it  is not modified by the mass deformation \cite{Y_mass_Liouville}. 
	The action \eqref{deformed_Liouville} defines  a continuous family of world-sheet CFT's to be used for the string quantization. They all have the central charge $c_L=3(1+Q^2)= 9$ (for our choice $Q=\sqrt{2}$) and the family is
	parametrized by the mass parameter  $\mathcal{M}$ of marginal deformation.

	The world-sheet action \eqref{deformed_Liouville} describes a nontrivial interacting world-sheet theory, and in order to extract information about the hadron spectrum of the 4D SQCD, we will rather use some indirect duality arguments.

	\subsection{$T$-duality}
	\label{sec:T-duality}

\begin{figure}[h]
		\centering
		\includegraphics[width=0.7\linewidth]{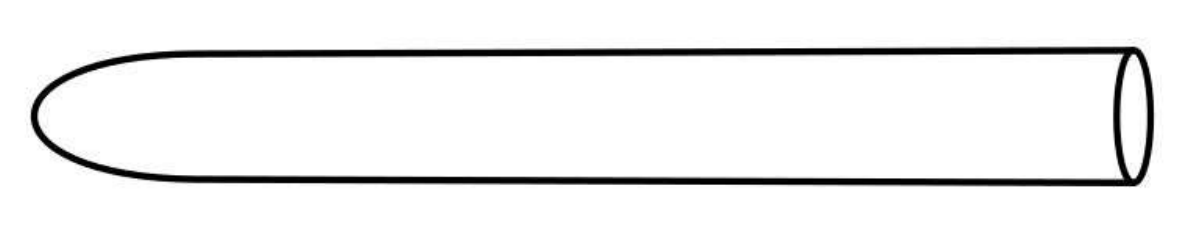}
		\caption{Cigar geometry of 2D black hole.}
		\label{fi:cigar}
	\end{figure}

	Now let us now show that the mass-deformed theory is actually $T$-dual to the 2D black hole.
	Consider the \ntwo  supersymmetric version of  the two-dimensional 
	black hole \cite{Wbh}, which is the SL($2, \mathbb{R}$)/U(1) coset WZNW theory 
	\cite{GVafa,GivKut,MukVafa,OoguriVafa95} 
	with the level
	$k$ of supersymmetric Ka\v{c}-Moody algebra.
	The bosonic part of the action reads
	\beq
	S_{\rm BH} = \frac{k}{4\pi}\int d^2 x \sqrt{h} \left\{(\pt_{\alpha} \rho)^2 +\tanh^2{\rho} \, (\pt_{\alpha}\theta)^2 \right\}
	+ \frac{1}{4\pi}\int d^2 x\sqrt{h}\, \Phi(\rho) R^{(2)}
	\label{BH}
	\eeq
	with the dilaton given by
	\beq
	\Phi(\rho)= \Phi_0 -\log{\cosh{\rho}}
	\label{dilaton_cigar}
	\eeq 
	so that the target 
	space has the form of a semi-infinite cigar with radial coordinate $0\leq \rho < \infty$ and angular coordinate $\theta\sim\theta+2\pi$, see Fig.~\ref{fi:cigar}. Here $k$ is related to $Q$ via \eqref{k_Q}.
	
	The first observation is that 
	the action \eqref{BH} can be rewritten as  the following  2D sigma model with a complex~\footnote{We are grateful to A.~Litvinov for important comments on this point.} metric \cite{ES}:
	\beq
	\label{Epert}
	S = \frac1{8\pi} \int d^2x \sqrt{h}\left( (\d_\mu\phi)^2 + (\d_\mu X)^2 -QR^{(2)}\phi +  e^{-Q(\phi -\phi_0)}\left(\d_\mu \phi - i\d_\mu X\right)^2\right)
	\eeq 
	Indeed, under the substitution
	\beq 
	\phi = \frac2{Q}\log\cosh \rho + \phi_0,\ \ \ X=\frac2{Q}\left(\theta 
	+ i\log\tanh \rho\right)
	\eeq 
	the action \eqref{Epert} turns into \eqref{BH} if one takes
	\beq
	\Phi_0=-\frac{Q}{2}\phi_0
	\label{Phi_0}
	\eeq
	in \eqref{dilaton_cigar}.
	Let us now make a $T$-duality transformation of the action \eqref{Epert}, rewriting it first as
	\begin{equation}
		\begin{aligned}
			S 
			&= \frac1{8\pi} \int d^2z \sqrt{g} \Big\{\left(1+e^{-Q\phi}\right)  (\d_\mu\phi)^2 + \left(1-e^{-Q\phi}\right) (\d_\mu X)^2 \\
			&\phantom{xxxxxxxxxxxxxxxxxxxxxxxxxxx} -2i e^{-Q\phi}\d_\mu \phi \d_\mu X - QR^{(2)}\phi 	\Big\} \\
			&\equiv  \frac1{4\pi} \int d^2z \sqrt{g} \Bigg\{\frac12\left[G_{\phi\phi}(\d_\mu\phi)^2 + G_{XX}(\d_\mu X)^2 +2G_{\phi X}\d_\mu \phi \d_\mu X\right] \\
			&\phantom{xxxxxxxxxxxxxxxxxxxxxxxxxxxxxxxxxxxxxxxx} - \frac{Q}2 R^{(2)}\phi \Bigg\}
		\end{aligned}
		\label{EL}
	\end{equation}
	%
	%\beqn 
	%%\label{EL}
	%&&S =  \frac1{8\pi} \int d^2z \sqrt{g}\left\{\left(1+e^{-Q\phi}\right) (\d_\mu\phi)^2 + \left(1-e^{-Q\phi}\right) (\d_\mu X)^2 -2i e^{-Q\phi}\d_\mu \phi \d_\mu X - QR^{(2)}\phi 
	%\right\} \equiv
	%\nonumber\\
	%&&\equiv  \frac1{4\pi} \int d^2z \sqrt{g}\left\{\frac12\left[G_{\phi\phi}(\d_\mu\phi)^2 + G_{XX}(\d_\mu X)^2 +2G_{\phi X}\d_\mu \phi \d_\mu X\right] - \frac{Q}2 R^{(2)}\phi 
	%\right\}
	%\eeqn 
	with $\phi_0=0$ for simplicity. By a standard trick
	\beq
	\label{tS}
	\hat{S} = \frac1{8\pi} \int d^2z \sqrt{g}\left( (\d_\mu\phi)^2 + V_\mu^2 -QR^{(2)}\phi + e^{-Q\phi}\left(\d_\mu \phi - iV_\mu \right)^2 + 2Y\epsilon_{\mu\nu}\d_\mu V_\nu\right)  
	\eeq 
	with Lagrange multiplier \(Y\) so that \(\left.\tilde{S}\right|_{\delta\tilde{S}/\delta Y=0}=S\). However, solving the equations \(\delta\tilde{S}/\delta V_\mu=0\) one finds
	\beq
	V_\mu =  \frac{ie^{-Q\phi}}{1-e^{-Q\phi}}\d_\mu \phi - \frac{\epsilon_{\mu\lambda}}{1-e^{-Q\phi}}\d_\lambda Y
	=  - \frac{G_{\phi X}}{G_{XX}}\d_\mu \phi - \frac{\epsilon_{\mu\lambda}}{G_{XX}}\d_\lambda Y
	\eeq 
	Substituting this back into \eqref{tS} we come to the dual action,
	\beqn
	&&S_{\rm dual}[\phi,Y] =  \frac1{4\pi} \int d^2z \sqrt{g}\left\{\frac12\left[\tilde{G}_{\phi\phi}(\d_\mu\phi)^2 + \tilde{G}_{YY}(\d_\mu Y)^2
	\right. \right.
	\nonumber\\
	&&
	\left. \left.
	+ 2i \tilde{B}_{Y\phi } \,\varepsilon_{\mu\nu}\pt_{\mu} Y \pt_{\nu}\phi\right] + R^{(2)}\Phi \right\} 
	\label{Sdual}
	\eeqn 
	with
	\beq
	\tilde{G}_{YY} = \frac1{G_{XX}},\ \ \ \ 
	\tilde{G}_{\phi\phi} = G_{\phi\phi} - \frac{G_{\phi X}^2}{G_{XX}}, \quad
	\tilde{B}_{ Y\phi} =\frac{G_{X\phi}}{G_{XX}}
	\eeq 
	and
	\beq
	\Phi = -\frac{Q}2\phi - \frac12\log G_{XX}
	\eeq 
	coming from changing integration measure \(DX \to DV\) (see e.g., \cite{Tdual}). We get, therefore, after the duality transformation,\footnote{An extra $B$-field \(\tilde{B}_{Y\phi} = i\frac{e^{-Q(\phi-\phi_0)}}{1-e^{-Q(\phi-\phi_0)}}\) in the  dual theory is almost inessential in two dimensions and we neglect it below.}
	\beqn
	&&\tilde{G}_{YY}(\phi) =  \tilde{G}_{\phi\phi}(\phi) = \frac1{1-e^{-Q(\phi-\phi_0)}}=g(\phi),
	\nonumber\\
	&&
	\Phi(\phi) = -\frac{Q}2\phi - \frac12\log(1-e^{-Q(\phi-\phi_0)}), 
	\label{Ybckg}
	\eeqn 
	coinciding exactly, after restoring the dependence on $\phi_0$ in the warp factor,
	with solution \eqref{g} for the metric and the dilaton \eqref{Phi_sol}.

	\subsection{Trumpet geometry and fluctuations}
	\label{sec:trumpet}
	
\begin{figure}[h]
		\centering
		\includegraphics[width=0.7\linewidth]{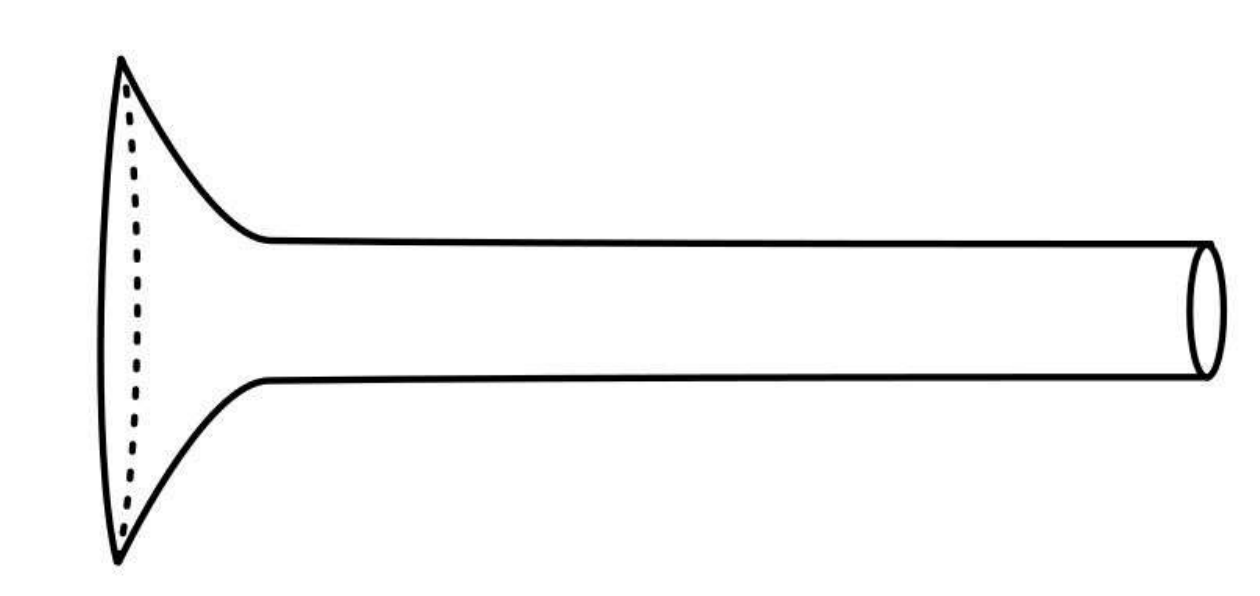}
		\caption{Trumpet geometry. Asymptotically, at \(\rho\to\infty\), it turns into a cylinder of radius \(Q\).}
		\label{fi:trumpet}
	\end{figure}

	Now let us make the world-sheet geometry of the mass-deformed \ntwo theory 
	\beq
	S_{\rm ws} =\frac{1}{4\pi}\int d^2 x \sqrt{h}\; \left\{\frac12\,g(\phi)\left[ (\pt_{\alpha}\phi)^2 + (\pt_{\alpha}Y)^2\right] 
	+ \;\Phi(\phi)\,R^{(2)} \right\},
	\label{mass_Liouville_no_sup}
	\eeq
	with the metric warp factor \eqref{g} and the dilaton \eqref{Phi_sol} more transparent. At $\phi=\phi_0$, this metric develops a naked singularity, so the geometry is defined at $\phi\geq \phi_0$.
	Making a change back to the radial coordinate,
	\beq
	e^{\frac{Q}{2}(\phi-\phi_0)}=\cosh{\rho},
	\label{rho}
	\eeq
	the action \eqref{mass_Liouville_no_sup} turns into
	\begin{equation}
	S_{\rm ws} =\frac{k}{4\pi}\int d^2 x \sqrt{h}\; \left( (\pt_{\alpha}\rho)^2 + \coth^2{\rho}\,(\pt_{\alpha}\vartheta)^2\right)
	+ \frac{1}{4\pi}\int d^2 x \sqrt{h}\Phi (\rho)\,R^{(2)} ,
	\label{trumpet}
	\end{equation}
	with \(\vartheta = \frac{Q}{2}\,Y\), 
	\beq
	\vartheta \sim \vartheta + 2\pi /k, 
	\label{Y_t_period}
	\eeq
	and the dilaton is now given by 
	\beq
	\Phi(\rho) = -\frac{Q}{2}\phi_0 -\log{(\sinh{\rho})}
	\label{dilaton_rho}
	\eeq
	The target manifold  of the dual theory \eqref{trumpet} looks like a ``trumpet''; see Fig.~\ref{fi:trumpet}. 
	This   metric   was first discussed in \cite{Giveon} and then in \cite{DijVerVer}  for the black hole  model \eqref{BH}. 
	The point is that the effective theory for the winding modes of the string in
	the cigar metric  \eqref{BH} of the 2D black hole is given by fluctuations of ``particles'' (momentum modes) in the  
	trumpet metric in \eqref{trumpet} \cite{DijVerVer} (see also  \cite{Mertens} for a review of the supersymmetric  version). It is just  a result of
	\emph{T}-duality transformation, which here takes the form
	\beq
	\sqrt{2k}\,  \tanh{\rho} \to \alpha' \; \frac{\coth{\rho}}{{\sqrt{2k}}}, \qquad  \alpha'=2.
	\label{tanh_to_coth}
	\eeq
	Taking the limit \(\rho\to\infty\), we see that \(k=1\) corresponds to the self-dual point [this is also seen, of course, from \eqref{m}]. 
	
	Let us now say a few words about the expected spectrum for the string in trumpet geometry. The  primary operators of the coset theory, corresponding to the string in the 2D black hole, are actually given at large $\rho$ by the  formula similar to the one in \eqref{genericj} (written for Liouville theory)
	\begin{equation}
		T_{j, m_L, m_R}
		\simeq
		%	\frac{1}{2j+1}
		e^{2i\,(m_L \theta_L + m_R  \theta_R)}\left[e^{2j \rho}+ R(j, m_{L,R} ; k) e^{-2(j+1) \rho}+\cdots\right],
		\label{genericj_cigar}
	\end{equation}
	where now the Liouville momenta and windings change roles and are interpreted as windings and momenta of the string in the cigar geometry, $n_2$ and $n_1$ respectively; see \eqref{m}. 
	%However, after making a T-duality transformation, their roles are exchanged again. 
	Hence, once we are interested in the operators \eqref{genericj_cigar} with $n_2\neq 0$ (\(m_L=-m_R\)), they correspond to the windings on the black hole side, and (in the minisuperspace approximation) it is natural to look for them as solutions of the target-space equations of motion in the trumpet geometry \cite{DijVerVer}. This target-space approximation works for large \(k\to\infty\), in contrast to direct Liouville quasiclassics, which works at large \(Q^2=2/k\). One can even be convinced that such reasoning is valid in the theory when both marginal deformations --- the Liouville superpotential and the trumpet warp factor --- are switched on independently.

	\subsection{Switching on both superpotential and mass deformations}
	\label{sec:superpot+mass}

	As is commonly believed, \ntwo Liouville theory with Liouville superpotential has a mirror description \cite{HoriKapustin} in terms of the \ntwo  supersymmetric version of  the two-dimensional  black hole \eqref{BH} with the level $k =\frac{2}{Q^2}$.
	The constant $\Phi_0$ entering the expression for the dilaton \eqref{dilaton_cigar} in this background is given by
	\beq
	\Phi_0 = \Phi_0^{(b)}= -\frac1k\,\ln{|b|}
	\label{Phi_0_b}
	\eeq
	in terms of the coefficient  $b$ (modulus of the conifold complex structure) in front of the Liouville superpotential by the  standard argument with the shift \(\phi\to\phi-Q\log|b|\equiv \phi -\phi_{\text{wall}}\), where $\phi_{\text{wall}}$ can be interpreted as a \textquote{position} of the ''Liouville  wall'' (superpotential).
	
	As was shown in \cite{Wbh}, the value of the dilaton at the horizon $\Phi_0$ determines the ADM mass of the black hole,
	\beq
	M_{BH} =  \frac{Q}{2}\, e^{-2\Phi_0} =\frac{1}{ \sqrt{2k}}\, e^{-2\Phi_0},
	\label{M_BH}
	\eeq
	and therefore the entropy
	\begin{equation}
		\label{dil_ent}
		S_{BH}  = \frac{M_{BH} }{T} = 2\pi e^{-2\Phi_0} 
	\end{equation}
	of the 2D black hole with the Hawking temperature \(T=\frac1{2\pi R}=\frac{Q}{4\pi}\). Hence, we see that the black hole entropy (and therefore the mass at fixed radius \(R=\sqrt{2k}\)) is completely determined by a single free parameter \(\Phi_0\) of the solution \eqref{dilaton_cigar}.
	
	For the supersymmetric black hole  associated with the Liouville theory, this parameter is totally determined \eqref{Phi_0_b} by the constant in front of the Liouville superpotential.
	Thus, for the black hole mass associated with the Liouville theory with superpotential deformation we have 
	\beq
	M_{BH}^{(b)} = \frac{Q}{2}\, |b|^{2/k}.
	\label{M_BH_b}
	\eeq
	On the other hand, as was shown in the previous section, the mass-deformed Liouville theory \eqref{mass_Liouville_no_sup} is \emph{T}-dual to \emph{the same} theory \eqref{BH} with the cigar geometry, where now the constant  part of the dilaton 
	\beq
	\Phi_0^{(\mathcal{M})}= -\frac{Q}{2}\,\phi_0=-\frac12\,\ln{\frac{|b|^2}{|\mathcal{M}|^2}},
	\label{Phi_0_M}
	\eeq
	which is given by the location of the singularity at $\phi =\phi_0$ of the warp factor \eqref{g}  of the metric, i.e., is actually determined by the mass deformation parameter $\mathcal{M}$.
	The associated black hole mass is
	\beq
	M_{BH}^{(\mathcal{M})} = \frac{Q}{2}\, \frac{|b|^2}{|\mathcal{M}|^2}.
	\label{M_BH_M}
	\eeq
	Let us now make a physically plausible conjecture.  If both marginal deformations are switched on [see \eqref{deformed_Liouville}],  one still gets the same dual black hole theory \eqref{BH} with cigar geometry, and we only have to fix the value of ``total'' \(\Phi_0\) or total mass of the black hole [they are related by \eqref{M_BH}] in terms of the initial data in the theory.  We conjecture that
	the total mass of the black hole upon switching on both deformations is the sum of masses \eqref{M_BH_b} and \eqref{M_BH_M},
	\begin{equation}
		M_{BH}^{\text{total}}= M_{BH}^{(b)}+ M_{BH}^{(\mathcal{M})}. 
		\label{phi_total}
	\end{equation}
	
	This gives for  the total mass and entropy of the black hole, corresponding to the theory with both deformations,
	\beq
	\frac{M_{BH}^{\text{total}}}{Q/2}= \frac{S_{BH}^{\text{total}}}{2\pi}=
	|b|^{2/k} +  \frac{|b|^2}{|\mathcal{M}|^2}.
	\label{M_BH_total}
	\eeq
	Let us now try to justify the conjecture \eqref{phi_total}, \eqref{M_BH_total}:
	\begin{itemize}
		\item Each term in the rhs \eqref{phi_total} dominates at large negative values of \eqref{Phi_0_b} or \eqref{Phi_0_M}, when $e^{-2\Phi_0} = e^{-2\Phi_0^{(b)}} + e^{-2\Phi_0^{(\mathcal{M})}}$ can be rewritten as 
		\begin{equation}
			e^{2|\Phi_0|} = e^{2|\Phi_0^{(b)}|} + e^{2|\Phi_0^{(\mathcal{M})}|} 
			\label{phi_total_mod}
		\end{equation}
		or, in the limit, as a ``tropical sum'', 
		\begin{equation}
			|\Phi_0| = |\Phi_0^{(b)}|\oplus|\Phi_0^{(\mathcal{M})}| = \max(|\Phi_0^{(b)}|,|\Phi_0^{(\mathcal{M})}| )
			\label{phi_tro}
		\end{equation}
		\item The asymptotic condition \eqref{phi_tro} does not uniquely fix the exact form of the function in \eqref{phi_total}, but it is reasonable to have a linear addition law for the entropy and mass in \eqref{M_BH_total}.
		\item At large $\mathcal{M}\to\infty$ the black hole mass does not feel the mass deformation. However, as we reduce the mass parameter $\mathcal{M}$, both terms in \eqref{M_BH_total} become of the same order and the system goes through a crossover. At small $\mathcal{M}$ the 
		second term in \eqref{M_BH_total} dominates.
		%\item In the ``flat region'' when \(\phi\to\infty\) both  \eqref{Phi_0_b} and \eqref{Phi_0_M} can be thought of as ``small'', and then \eqref{phi_total} gives approximately \(\Phi_0 \approx \Phi_0^{(b)}+\Phi_0^{(\mu)}\) usual sum (instead of tropical), as naively follows from the linearized world-sheet action \eqref{deformed_Liouville}.
	\end{itemize}
	Formula \eqref{phi_total} is one of the key points of our paper. It comes out of the fact that both (undeformed and mass-deformed) theory are formulated in the same string background, and the only parameter which can depend on the mass deformation is the dilaton constant $\Phi_0$. The exact form of this formula can be verified only by rather complicated analysis of the world-sheet theory \eqref{deformed_Liouville}, so we proposed instead its most straightforward form motivated by the idea that the full mass or entropy of the black hole can be naturally thought to be the sum of two independent contributions associated with the Liouville superpotential and the mass deformation. It obviously reproduces both \textquote{boundary} cases when either the Liouville superpotential deformation or the nontrivial metric warp factor, arising due to the mass deformation, dominates. These two limits of cigar theory are effectively described either by its mirror or $T$-dual formulations, which relate  \eqref{deformed_Liouville} to the black hole  \eqref{BH} with $\Phi_0$ determined either by \eqref{Phi_0_b} or by \eqref{Phi_0_M} respectively.

	\section {String spectrum after mass deformation}
	\label{sec:hyper}
	\setcounter{equation}{0}
	
	In this section, we use 
	an effective gravity approach  to study the string spectrum  associated with the mass-deformed \ntwo Liouville world sheet theory. First, we review 
	the setup for this problem developed in \cite{Y_mass_Liouville},  then we study the associated Schr\"odinger equation and find its discrete  spectrum. We show that this spectrum coincides with the spectrum of a 2D black hole with cigar geometry, confirming the result of the previous section that these theories are related by $T$-duality. %{\color{red}Not sure that this should be deleted: To do so
		%in this section we switch off the Liouville superpotential in \eqref{deformed_Liouville} 
		%considering only the mass deformation of the Liouville theory}.

	\subsection{Tachyon equations}
	\label{sec:tachyon_eq}
	
	Primary tachyon vertex operators \eqref{genericj} can be described as scalar fields in the effective supergravity \eqref{gravity_action}. To take them into account, we add the tachyonic (with the \textquote{wrong} sign in the D-dimensional space-time) term,
	\beq
	S_{\rm tachyon}= \frac1{2\kappa^2}\int d^D x \, \sqrt{-G}\,e^{-2\Phi}\,\left\{-G^{MN}\pt_M \bar{\mathcal{T}}_{j,m}\pt_N \mathcal{T}_{j,m} 
	+ |\mathcal{T}_{j,m}|^2\right\}.
	\label{tachyon_action}
	\eeq
	to the gravity action \eqref{gravity_action}, cf. \cite{Polch_90}. 
	This gives the equation for the tachyon field,
	\beq
	D_N D^N \mathcal{T}_{j,m} -2\pt_N\Phi\pt^N \mathcal{T}_{j,m} + \mathcal{T}_{j,m}=0
	\label{T_eq_gen}
	\eeq
	and we neglect the back reaction of tachyons on the background \eqref{g}, \eqref{Phi_sol} for the metric and the dilaton.
	
	Dressing the tachyon with the dependence on the 4D coordinates as in \eqref{dressed_tachyon}
	we rewrite the tachyon equation \eqref{T_eq_gen} in the form,
	\beqn
	&&D_n D^n T_{j,m} -2\pt_n\Phi\pt^n T_{j,m}
	+ \left(1+ M_T^2\right)T_{j,m}  =
	\nonumber\\
	&&
	= D_n D^n T_{j,m} -2\pt_n\Phi\pt^n T_{j,m}
	+ 2\Delta_{j,m}T_{j,m}
	=0,
	\label{T_eq}
	\eeqn 
	where derivatives  are now taken only with respect to two internal coordinates,  $\{\phi,Y\}$. In \eqref{T_eq} we have used the relation,
	\beq
	M_T^2 = 2\Delta_{j,m} -1,
	\label{M_4D}
	\eeq
	which coincides with Eq. \eqref{tachphys},
	obtained by world sheet methods, see \cite{SYlittles} for details. Here  the conformal dimension $\Delta_{j,m}$ is given by \eqref{dimV}. The relation\eqref{M_4D} is also easily verified by substitution,
	\beq
	T_{j,m} \approx e^{Q[j\phi +imY]},
	\label{T_0}
	\eeq
	in the flat background with the linear dilaton \eqref{linear_dilaton}.
	
	Since the warp factor \eqref{g} does not depend  on $Y$, one can look for solutions of \eqref{T_eq} using the ansatz, 
	\beq
	T_{j,m}(\phi,Y) = e^{\Phi}\, e^{iQmY}\Psi_{j,m} (\phi),
	\label{T_Psi}
	\eeq
	where the factor $e^{\Phi}$ kills first derivatives in \eqref{T_eq}.
	This substitution gives the Schr\"odinger equation for the wave function $\Psi_{j,m}(\phi)$
	\cite{Y_mass_Liouville},
	\beq
	-\pt_{\phi}^2 \Psi_{j,m} + V_{{\rm eff}} (\phi)\Psi_{j,m} = E_{j} \Psi_{j,m},
	\label{Psi_eq}
	\eeq
	where the potential is given by
	\beq
	V_{{\rm eff}} (\phi) =  - \frac{Q^2}4\frac1{\left(e^{Q(\phi-\phi_0)}-1\right)^2} - \frac{Q^2(m^2-j(j+1))}{e^{Q(\phi-\phi_0)}-1}\ \stackreb{\phi\to \phi_0}{\sim} - \frac1{4(\phi-\phi_0)^2} + \ldots
	\label{pot}
	\eeq
	while energy levels are determined by $j$,
	\beq
	E_j =-Q^2\left(j+\frac12\right)^2.
	\label{E}
	\eeq
	The potential \eqref{pot} is attractive for $m^2-j(j+1) >0$ and tends to zero at $\phi\to\infty$. 
	Therefore, one  expects the  continuous spectrum~\footnote{Note that we are looking for the spectrum of normalizable and borderline-normalizable states. The continuous spectrum mentioned here corresponds to the {\em principal} continuous spectrum of the cigar theory \cite{MukVafa,DixonPeskinLy,Petrop,Hwang,EGPerry}, see \cite{EGPerry-rev} for a review.} with $j=-\frac12 + i\mathbb{R}$ with positive $E_j$ and a  discrete spectrum with negative $E_j$. 
	
	Equation \eqref{T_eq} for tachyons can also be written in the trumpet background \eqref{Ybckg}. Introducing the wave function via
	\beq
	T_{j,m}= \frac{e^{2im\vartheta}}{\sqrt{\sinh{2\rho}}}\,\Psi^{(t)}_{j,m}  \,,
	\label{T_Psi_t}
	\eeq
	[cf. with \eqref{T_Psi}] one gets a similar to \eqref{Psi_eq}  Schr\"odinger equation, 
	\beq
	-\pt_{\rho}^2 \Psi^{(t)}_{j,m} + V_{{\rm eff}} (\rho)\Psi^{(t)}_{j,m} 
	= E^{(t)}_{j} \Psi^{(t)}_{j,m} \,,
	\label{Psi_eq_t}
	\eeq
	where the effective potential is now given by
	\beq
	V_{{\rm eff}} (\rho) =  - \frac1{4\sinh^2\rho} + \frac{1-16m^2}{4\cosh^2\rho}
	%-\frac{1}{\sinh^2{2\rho}} +4m^2\left(\tanh^2{\rho}-1\right).
	\label{pot_t}
	\eeq
	and  the  energy levels are just
	\beq
	E^{(t)}_j =-\left(2j+1\right)^2.
	\label{E_t}
	\eeq
	The wave function $\Psi^{(t)}_{j,m}$ at large $\rho$, where the potential goes to zero 
	has the form
	\beq
	\Psi^{(t)}_{j,m}(\rho) \approx e^{2\left(j+\frac12\right)\rho}
	\label{Psi_t}
	\eeq
	The potential  \eqref{pot_t} is also attractive and, of course, gives the same spectrum as 
	the one from \eqref{pot}. Solutions of Eqs. \eqref{Psi_eq_t} and \eqref{Psi_eq} are related by
	\begin{equation}
		\Psi^{(t)}_{j,m}(\rho)= \Psi_{j,m}(\phi)\sqrt{\coth\rho}
	\end{equation}
	together with \eqref{rho}.
	
	Equation \eqref{Psi_eq_t} describes the momentum modes with $n_2\neq 0$ [see \eqref{m}]
	in the trumpet geometry \eqref{trumpet}. The nonsupersymmetric version of this equation was considered first in  \cite{DijVerVer} for winding modes (also with $n_2\neq 0$) for the dual black hole theory with the cigar geometry.

	\subsection{Equations for massive ``gravitons''}
	\label{sec:gravitons}
	
	Consider now string states with integer $m$ and $j$ associated with massive ``gravitons''; see \eqref{graviton}. The gravity action for the scalar components $V_{j,m}$ of the corresponding spin-2 supermultiplets has the form
	\beq
	S_{\rm graviton}= \frac1{2\kappa^2}\int d^D x \, \sqrt{-G}\,e^{-2\Phi}\,\left\{-G^{MN}\pt_M \bar{V}_{j,m}\pt_N V_{j,m} \right\},
	\label{graviton_action}
	\eeq
	where the mass term is absent in contrast to \eqref{tachyon_action}.
	Much in the same way as before for the tachyons, one gets the equations of motion 
	\beq
	D_n D^n V_{j,m} -2\pt_n\Phi\pt^n V_{j,m}
	+  M_V^2V_{j,m} =0,
	\label{V_eq_M_4D}
	\eeq
	where $V_{j,m}$ depends only on $\phi$ and $Y$. Taking the limit of large $\phi$, 
	where $g(\phi)\to 1$, and looking for a solution of \eqref{V_eq_M_4D} with the ansatz
	\beq
	V_{j,m} \approx e^{Q[j\phi +imY]}.
	\label{V_0}
	\eeq
	we get
	\beq
	M_V^2 = 2\Delta_{j,m}.
	\label{M_4D_g}
	\eeq
	This coincides with \eqref{gravitonmass} for $j=-1$. Substituting this into \eqref{V_eq_M_4D}, we get
	\beq
	D_n D^n V_{j,m} -2\pt_n\Phi\pt^n V_{j,m}
	+  2\Delta_{j,m} V_{j,m} =0,
	\label{V_eq}
	\eeq
	This equation has the same form as the equation of motion \eqref{T_eq} for tachyons; thus we conclude that the wave function for massive ``gravitons'' defined via
	\beq
	V_{j,m}(\phi,Y) = e^{\Phi}\, e^{iQmY}\Psi_{j,m} (\phi)
	\eeq
	satisfies the same Schr\"odinger equation \eqref{Psi_eq} or \eqref{Psi_eq_t}, and we need only select solutions with different (integer) values of $m$ and $j$.

	\subsection{Exact spectrum and profiles}
	\label{sec:hypergeometry}
	
	Luckily enough, the second-order differential equation \eqref{Psi_eq} can be solved exactly, and the solution profiles are even expressible in terms of elementary functions. The conditions for the spectrum that arise in the process will also give us an important hint.
	
	We start with a substitution 
	\beq
	\Psi(x)\equiv w(x)\frac{\sqrt{e^{Qx}-1}}{e^{Qmx}}
	\label{subst}
	\eeq
	and change  the variable $x\equiv\phi-\phi_0$ to  $z\equiv 1-e^{Qx}$.  Then $w(z)$ satisfies the hypergeometric equation 
	\begin{equation}
		w''+ \frac{1-2(1-m)z}{z(1-z)}w'+\frac{(j+m)(j+1-m)}{z(1-z)}w=0,
		\label{w_eq_z} 
	\end{equation}
	and the solution regular at $z=0$ (i.e. at $\phi = \phi_0$) is given by
	\begin{equation}
		w_{reg}=F(\alpha,\beta,1 ,z),
		\label{wreg_z1}
	\end{equation}
	a standard hypergeometric function with
	\begin{equation}
		\alpha=-j-m,\quad \beta=1+j-m.
		\label{alpha_beta}
	\end{equation}
	%
	%
	%The second linear independent  solution, singular at $z=0$ has the form
	%\begin{equation}
	%		w_{sing}=F(\alpha,\beta,1,z)\ln z+\sum_{k=1}^{\infty}(z)^k\frac{(\alpha)_k(\beta)_k}{(k!)^2}\{\psi(\alpha+k)-\psi(\alpha)+\psi(\beta+k)-\psi(\beta)-2\psi(k+1)+2\psi(1)\}, 
	%\end{equation}
	%where $(a)_n=a(a+1)...(a+n-1)$ is the Pochhammer symbol.
	%
	The second (linearly independent) solution is singular at $z=0$, nonsymmetric under $m\leftrightarrow-m$, and therefore should be discarded (see Appendix~\ref{app:sing_cont}).
	
	To continue analytically the solution \eqref{wreg_z1} to the region $z\to-\infty$ along the negative real axis (corresponding to $\phi\to\infty$) one can use
	(9.132) from \cite{Ryzhik} and get
	\begin{equation}
		\begin{aligned}
			w_{reg}\bigg|_{z\to-\infty}
			&=F(\alpha,\beta,1;z)\bigg|_{z\to-\infty} \\
			&=\frac{\Gamma(1+2j)}{\Gamma(1+j-m)\Gamma(1+j+m)}(-z)^m
			\bigg[(-z)^j+R_{reg}
			(-z)^{-1-j}+\cdots\bigg].
		\end{aligned}
		\label{F_z_infty}
	\end{equation}
	At $\phi \to +\infty$ ($z\to-\infty$), for the function $\Psi_{j,m}\sim (-z)^{-m+1/2}w_{reg}(z)$
	so the first term in the brackets corresponds to the falling exponent $e^{Q(j+1/2)\phi}$, while $(-z)^{-1-j}$ gives the rising $e^{-Q(j+1/2)\phi}$ in the wave function $\Psi_{j,m}$ corresponding to \eqref{genericj} for  $j\leq -1/2$. The reflection coefficient in \eqref{F_z_infty} is given by
	\begin{equation}
		R_{reg}=\frac{\Gamma(-1-2j)\Gamma(1+j-m)\Gamma(1+j+m)}{\Gamma(1+2j)\Gamma(-j-m)\Gamma(-j+m)}.
		\label{refl_reg}
	\end{equation}
	The discrete spectrum is determined by the zeros of the reflection coefficient, see Sec.~\ref{sec:vertices},
	which ensures that the rising exponent is absent at
	\beq
	-j-|m|=-n, \qquad n=0,1,...,
	\label{discrete_condition}
	\eeq 
	The reflection coefficient \eqref{refl_reg} coincides with the $k$-independent part of the exact formula \eqref{refl_GK}, obtained using the $SL(2,R)/U(1)$ coset 
	(super)conformal theory, and matches with the statement of the previous section that
	the mass-deformed \ntwo Liouville theory is $T$-dual to the supersymmetric black hole with cigar
	geometry (so these theories have the same spectrum of string states). Indeed, we see that Eqs. \eqref{w_eq_z} and \eqref{Psi_eq_t} do not depend on $k$, so our result \eqref{refl_reg} from the effective gravity approach reproduces  the exact expression \eqref{refl_GK} only in the limit of large $k$ (or the  leading order in $\alpha'$), while missing the $\alpha' \sim 1/k$ world-sheet corrections~\footnote{Although the target space geometry \eqref{deformed_Liouville} itself in the supersymmetric case (in contrast to the bosonic 2D black hole) does not receive the $\alpha' \sim 1/k$ corrections.} (even in the supersymmetric case). We point out again here, that  although the gravity approximation in the mirror to Liouville ``cigar theory'' works only at large \(k\to\infty\), still the $k$-independent conclusions can be  applied to the most interesting critical \(k=1\) case. In particular, the reflection coefficient \eqref{refl_reg} reproduces the correct discrete
	spectrum \eqref{discrete} however misses  the lower unitarity bound for allowed values of $j$ in \eqref{no_ghosts}.
	
	Under conditions \eqref{discrete_condition} for integer and half-integer $j$ and $m$ the parameters \eqref{alpha_beta} of the hypergeometric equation \eqref{w_eq_z} become integers, and it can be solved in terms of rational functions. For the branch of solutions to \eqref{discrete_condition} with
	\(m\geq |j|>0\) one gets the polynomial solutions,
	\begin{equation}
		\begin{aligned}
			w_{j,m>0}^{(+)}(z)
			&= F(-j-m,1+j-m,1 ;z) \\
			&= \left.\frac{(1-z)^{2m}}{n!}\frac{d^n}{dz^n}\left(z^n(1-z)^{j-m}\right)\right|_{n=j+m}
		\end{aligned}
		\label{w_plus}
	\end{equation}
	For negative values of \(m\leq j<0\), using a Kummer identity one can write 
	\begin{equation}
		\begin{aligned}
			w_{j,m<0}^{(-)}(z)
			&= F(-j-m,1+j-m,1 ;z) \\
			&= (1-z)^{2m}F(-j+m,1+j+m,1 ;z) \\
			&= (1-z)^{-2|m|}F(-j-|m|,1+j-|m|,1 ;z) \\
			&= \left.\frac{1}{n!}\frac{d^n}{dz^n}\left(z^n(1-z)^{j+m}\right)\right|_{n=j-m}
		\end{aligned}
		\label{w_minus}
	\end{equation}
	%
	%\beqn
	%&& w_{j,m<0}^{(-)}(z)=F(-j-m,1+j-m,1 ;z) = (1-z)^{2m}F(-j+m,1+j+m,1 ;z) =
	%\nonumber\\
	%&&
	%=(1-z)^{-2|m|}F(-j-|m|,1+j-|m|,1 ;z) = \left.\frac{1}{n!}\frac{d^n}{dz^n}\left(z^n(1-z)^{j+m}\right)\right|_{n=j-m}
	%\label{w_minus}
	%\eeqn 
	where the bottom line follows from \eqref{w_plus}. We see that \(w_{j,m<0}^{(-)}(z) = w_{j,|m|}^{(+)}(z)(1-z)^{-2|m|}\), and therefore both degenerate hypergeometric  solutions with \(m=\pm|m|\) by \eqref{subst} give rise  to the same wave function \(\Psi_{j,\pm|m|}\).
	To conclude this section, we list  the wave functions $\Psi(x)$ for the low-lying string states
	with allowed values of $j=-1/2$ and $j=-1$, see the table below,
	\begin{center}
		\begin{tabular}{|c|c|c|c|c|c|c|} 
			\hline
			$j$ & $m$ & $\alpha$ & $\beta$ & $w_{reg}(z)$& $\Psi(x)$&Asympt. $x\rightarrow\infty$ \\ 
			\hline\hline
			\multirow{4}{*}{$-\frac{1}{2}$} & $-\frac{1}{2}$ & $1$ & $1$ & $\frac{1}{1-z}$& \multirow{2}{*}{$e^{-\frac{Qx}{2}}\sqrt{e^{Qx}-1}$}&\multirow{2}{*}{$1$} \\ 
			\cline{2-5}
			& $\frac{1}{2}$ & $0$ & $0$ & $1$& & \\ 
			\cline{2-7}
			& $-\frac{3}{2}$ & $2$ & $2$ & $\frac{1+z}{(1-z)^3}$& \multirow{2}{*}{$e^{-\frac{3Qx}{2}}(2-e^{Qx})\sqrt{e^{Qx}-1}$}&\multirow{2}{*}{$-1$} \\
			\cline{2-5}
			& $\frac{3}{2}$ & $-1$ & $-1$ & $1+z$&  &\\
			\hline
			\multirow{4}{*}{$-1$} & $-1$ & $2$ & $1$ & $\frac{1}{(1-z)^2}$& \multirow{2}{*}{$e^{-Qx}\sqrt{e^{Qx}-1}$} &\multirow{2}{*}{$e^{-\frac{Q}{2}x}$}\\
			\cline{2-5}
			& $1$ & $0$ & $1$ & $1$& & \\
			\cline{2-7}
			& $-2$ & $3$ & $2$ & $\frac{1+2z}{(1-z)^4}$&\multirow{2}{*}{ $e^{-2Qx}(3-2e^{Qx})\sqrt{e^{Qx}-1}$}&\multirow{2}{*}{$-2e^{-\frac{Q}{2}x}$} \\ 
			\cline{2-5}
			& $2$ & $-1$ & $-2$ & $1+2z$&  &\\ 
			\hline
		\end{tabular}
	\end{center}
	and for the ``border case'' \(j=-\frac{k}2\), \(m=\pm\frac{k}2\) these functions take the form
	\begin{equation}
		\Psi^{(t)}_{-k/2,\pm k/2}(\rho)=\sqrt{\tanh\rho}\cosh^{1-k}\rho,\ \ \ \  \Psi_{-k/2,\pm k/2}(\phi)=e^{-\frac{\phi-\phi_0}{Q}}\sqrt{e^{Q(\phi-\phi_0)}-1}
		\label{Psi_bk}
	\end{equation}

	\section{SQCD interpretation}
	\label{sec:SQCD_interpretation}
	\setcounter{equation}{0}
	
	The results of Sec.~\ref{sec:superpot+mass} show that when both the superpotential and the mass deformation are switched on in 
	the \ntwo Liouville theory, one gets the same cigar geometry in the dual black hole picture, and the only effect of the mass deformation is an increase of the black hole mass, see \eqref{M_BH_total}.  This means that  as we reduce the mass parameter $\mathcal{M}$
	the mass spectrum of the string states does not change. In particular, the 
	low-lying hadron spectrum of the mass-deformed 4D SQCD  is still given by Eqs.~\eqref{tachyonmass} and \eqref{gravitonmass} from Sec.~\ref{sec: 4D spectrum}.
	
	This rather surprising result from the string theory side seems to run into a contradiction with the intuitive expectations on the field theory side. One would expect the emergence of new hadronic states in SQCD with the masses determined by $\mathcal{M}$, which become lighter as we reduce $\mathcal{M}$. Now we are going to demonstrate that in \ntwo supersymmetric QCD this intuitive expectation 
	%might appear 
	is actually incorrect under rather mild assumptions.

	Of course, one cannot calculate the hadron masses directly from SQCD, since the theory is at an extremely strong-coupling regime $g^2\to \infty$ (associated with $\beta=0$), and therefore our approach is to use the effective string theory of the critical non-Abelian string to find the hadron spectrum in SQCD. 
	However, we can make certain plausible assumptions on the hadron mass dependence on quark mass parameters, typical for \ntwo SQCD, and on the quantum numbers of allowed states. In particular, we assume that the squared hadron masses, 
	depending on $\xi$ and  on quark masses, are given by the sum of the ($\xi$-dependent) non-BPS part and  the squared
	central charge  $Z_{BPS}(m_i, \tilde{m}_j)$ of the \ntwo supersymmetric algebra for a given state,
	\beq
	m_H^2 = m^2_{\text {non-BPS}}(\xi) + |Z_{BPS}(m_i, \tilde{m}_j)|^2.
	\label{BPS+non-BPS}
	\eeq
	The BPS central charges are given by exact formulas, depending on  the global symmetry charges of a given state, cf. \cite{SW2}.
	
	To illustrate \eqref{BPS+non-BPS}, consider the masses of the perturbative states (quarks and gluons and their superpartners) in the Higgs phase at weak coupling. The non-BPS part of the mass, say, of the squark $q^{kA}$ , $A=1,...,N_f$, arising due to squark condensation \eqref{qvev},
	is determined by the $D$-term in the SQCD scalar potential
	%\beq
	$\left(|q^A|^2- |\tilde{q}_A|^2-N\xi \right)^2$,
	%\eeq
	see \cite{SYrev} for details, and gives $m^2_{\text {non-BPS}}\sim g^2\xi $. On the other hand, the BPS part of the mass of  $q^{kA}$ arises from the $F$-term in the SQCD scalar potential,
	%\beq
	$\left|\left(a+m_A\right)q^A\right|^2$
	%\eeq
	which turns into $|m_k-m_A|^2$ when VEVs of adjoint scalars \eqref{avev} are taken into account.
	
	To discuss the BPS parts of the masses of 4D SQCD states, it is convenient to use the 2D-4D correspondence mentioned in Sec.~\ref{sec:wcp},
	which claims that the BPS spectrum  of 2D world sheet theory  coincides
	with the BPS spectrum of 4D SQCD in the quark vacuum \eqref{qvev} at $\xi=0$.
	For example, the quark $q^{kA}$ with $A=N+j$ has the same mass $|m_k-\tilde{m}_j|$ as  the gauge-invariant ``meson'' $w^{kj}=n^k\rho^j$ in the \wcpN model.

	%For example, the quark $q^{kA}$ with $A=N+j$ has the same mass $|m_k-\tilde{m}_j|$ as $\rho^j$ state in the vacuum where
	%$n^k$ classically develop VEV, see \eqref{wcpN_vac}  in Sec.~\ref{sec:wcp}. This state in \wcpN model can be described by the  gauge invariant ''meson'' $n^k\rho^j$.

	\subsection{BPS central charges}
	\label{sec:central_charges}

	We have already discussed  the global symmetry group  \eqref{c+f} of the $\mathbb{WCP}(N,N)$ model
	in Sec.~\ref{sec:wcp}, so that the $n^i$ and $\rho^j$ fields belong to its representations \eqref{n_rho_representations}.
	The exact formula for their BPS central charges has the form
	\begin{equation}
		\label{Zmq}
		Z_{BPS} = i \vec{m}_n \vec{q}_n - i \vec{m}_\rho \vec{q}_\rho 
	\end{equation}
	where \(\vec{m}_n = \{m_1,\ldots,m_N\}\) and  \(\vec{m}_\rho = \{\tilde{m}_1,\ldots,\tilde{m}_N\}\) are the masses of \(n\) and \(\rho\) fields (they coincide with the masses of $2N$ flavors of quarks, see Sec.~\ref{sec:wcp}). Here we use that
	\(\mathfrak{u}_N= \mathfrak{su}_N\oplus \mathfrak{u}_1\) and consider the \(\mathfrak{gl}_N\supset \mathfrak{u}_N\) charges  \(\vec{q}_n\) and
	\(\vec{q}_\rho\)  (instead of \(\mathfrak{sl}_N\supset \mathfrak{su}_N\) charges), since
	both $n^i$ and $\rho^j$ fields transform under the fundamental representation \(\mathbf{N}\) of \(\mathfrak{gl}_N\). 
	
	In the simplest ``canonical'' basis,	
	\begin{equation}
		(\vec{e}_i)_k\equiv\delta_{ik},\ \ \ \ i,k=1,\ldots,N
	\end{equation}
	the Cartan \(\mathfrak{gl}_N\)-generators
	\begin{equation}
		(H_i)_{jk}=\delta_{ij}\delta_{ik}
		\label{un_cartan}
	\end{equation}
	determine the charges $\vec{q}_{n^i}$ by
	\begin{equation}
		H_k\vec{e_i}=(q)_k\vec{e_i}
		\label{gln_charges}
	\end{equation}
	for both $\vec{q}_{n^i}$ and $\vec{q}_{\rho^j}$, and therefore  the ``perturbative states'' $n^i$ and $\rho^j$ have charges,
	\begin{equation}
		\vec{q}_{n^i}=\vec{e_i},\quad \vec{q}_{\rho^j}=\vec{e_j},
		\label{fund_gln}
	\end{equation}
	so that they form the basis of the lattice \(\mathbb{Z}^N\subset \mathbb{R}^N\). 
	Substituting \eqref{fund_gln} into the BPS formula \eqref{Zmq}, we immediately obtain, for an arbitrary ``mesonic'' $n^i\rho^j$ state,
	\begin{equation}
		Z_{n^i\rho^j}=i(m_i-\tilde{m}_j),
		\label{Z_pert}
	\end{equation}
	as expected. Any charge vector can be decomposed as
	\begin{equation}
		\vec{q} = \frac1N \vec{e} B + \vec{q}_\bot
	\end{equation}
	where
	\(\vec{e}=\sum_j \vec{e}_j\) is the diagonal vector, \(B=\sum_j q_j\) is the corresponding baryonic charge, while \(\vec{q}_\bot\) is the vector of the \(sl_N\) charges, 
	satisfying \(\vec{q}_\bot\cdot \vec{e}=0\).
	In the  \(N-1\)-dimensional hyperplane orthogonal to \(\vec{e}\), the basis vectors \(\{\vec{e}_i\}\) project to \(\{\vec{\nu}_i=\vec{e}_i-\frac1N \vec{e}\}\) and  satisfy
	\begin{equation}
		\vec{\nu}_i \vec{\nu}_j = \delta_{ij}-\frac1N,\ \ \ \sum_j \vec{\nu}_j = 0.
	\end{equation}
	For the set of \(\mathfrak{sl}_N\) charges,
	let us take   the root basis \(\vec{\alpha}_{ij} = \vec{\nu}_i -\vec{\nu}_j=\vec{e}_i-\vec{e}_j\) for the simple roots
	\begin{equation}
		\vec{\alpha}_{a} =\vec{\alpha}_{a,a+1},\ \ \ \ a=1,\ldots,N-1,\ \ \ \ \vec{\alpha}_{a}^2=2
		\label{above_sln_cartan}
	\end{equation}
	which have the charges  \(h_1 = \diag(1,-1,0,\ldots,0)\), \(h_2 = \diag(0,1,-1,0,\ldots,0)\), ..., \(h_{N-1}= \diag(0,\ldots,0,1,-1,)\) with respect to \(\mathfrak{sl}_N\) Cartan generators
	\begin{equation}
		h_a=H_a-H_{a+1},\ \ \ \ a=1,\ldots,N-1
	\end{equation}
	also satisfying
	\begin{equation}
		\label{Lie_slN}
		[h_a,e_b]=C_{ab}e_b,\ \ \ \ [h_a,f_b]=-C_{ab}f_b,\ \ \ \ [e_a,f_b]=\delta_{ab}h_a,\ \ \ \ \ a,b=1,\ldots,N-1
	\end{equation}
	with \(C_{ab}=\vec{\alpha}_{a}\cdot\vec{\alpha}_{b}=2\delta_{ab}-\delta_{a+1,b}-\delta_{a,b+1}\) being the Cartan matrix of  \(\mathfrak{sl}_N\), while \(e_a=E_{a,a+1}\) and \(f_a=E_{a+1,a}\) are the matrices of simple roots.
	The corresponding charges in the fundamental representation \(\mathbf{N}\),
	\begin{equation}
		h_a(\vec{\nu}_j ) = \vec{\alpha}_{a}\cdot \vec{\nu}_j = \delta_{aj}-\delta_{a+1,j}
	\end{equation}
	are always integers.
	One therefore gets
	\begin{equation}
		\vec{q}_\bot = \sum_{i=1}^N q_i  \vec{\nu}_i = \sum_{a=1}^{N-1} \mathsf{q}^a\vec{\alpha}_{a}
	\end{equation}
	so that 
	\begin{equation}
		\sum_{b=1}^{N-1} C_{ab}\mathsf{q}^b = \sum_{i=1}^N q_i  \vec{\alpha}_{a}\cdot\vec{\nu}_i =q_a-q_{a+1} \equiv Q_a,\quad a=1...N-1
		\label{Qa_def}
	\end{equation}
	is the charge with respect to the $a$th $\mathfrak{sl}_2$ subalgebra (if $Q_a=0$ for a certain $a$, the state is a singlet with respect to this  $\mathfrak{sl}_2$ subalgebra).
	
	Hence \(\mathsf{q}^a=\sum_{b=1}^{N-1} C^{-1}_{ab}Q_b\) are ``dual charges'' [not necessarily integers, in contrast to \(\{Q_b\}\), due to \(C^{-1}_{ab}=\min(a,b)-\frac{ab}{N}\)], and therefore
	\begin{equation}
		\begin{aligned}
			Z_{BPS} &= i \frac{B}N\left(\vec{e}\cdot \vec{m}_n -\vec{e} \cdot \vec{m}_\rho\right) +  i \vec{m}_n \vec{q}_n^\bot - i \vec{m}_\rho \vec{q}_\rho^\bot \\
			&=	i \frac{B}N\left(\sum_{j=1}^N m_j - \sum_{j=1}^N \tilde{m}_j\right) \\
			&\phantom{xxxxxxx} +   i \sum_{a=1}^{N-1}\left( (m_a-m_{a+1}) \mathsf{q}^a
			-   (\tilde{m}_a-\tilde{m}_{a+1}) \tilde{\mathsf{q}}^a\right) \\
			&=	i \frac{B}N\left(\sum_{j=1}^N m_j - \sum_{j=1}^N \tilde{m}_j\right) \\
			&\phantom{xxxxxxx} +   i \sum_{a,b=1}^{N-1}\left( (m_a-m_{a+1})C^{-1}_{ab}Q_b
			-   (\tilde{m}_a-\tilde{m}_{a+1})C^{-1}_{ab}\tilde{Q}_b\right)
		\end{aligned}
		\label{ZmqC}
	\end{equation}

	\subsubsection{\(SU(2)\) case}
	
	Now let us consider the starting point of our interpolation procedure at $\mathcal{M}\to\infty$, namely the \wcpt model. The global symmetry group \eqref{c+f} of this model is
	\beq
	SU(2)\times SU(2)\times U(1)_B,
	\label{glob_sym_N=2}
	\eeq
	and each SU(2) factor here is broken down to U(1) by the quark mass difference.
	
	Following \eqref{un_cartan}, \eqref{gln_charges}, and \eqref{fund_gln}, one gets
	\begin{equation}
		H_{1}=\text{diag}(1,0),\quad H_{2}=\text{diag}(0,1),
		\label{u2_gens}
	\end{equation}
	and 	
	\begin{equation}
		\vec{q}_{n^1}=\vec{q}_{\rho^1}=(1,0),\ \ \ \ \ \vec{q}_{n^2}=\vec{q}_{\rho^2}=(0,1)
		\label{u2_charges}
	\end{equation}
	while, in the $\mathfrak{su}_2\oplus \mathfrak{u}_1$ terms, for \(h=H_1-H_2\) and the unit \(H_1+H_2\), we get
	\begin{equation}
		\begin{aligned}
			&(Q,B)_{n^1}=(\tilde{Q},B)_{\rho^1}=\left(1,1\right),\\
			&(Q,B)_{n^2}=(\tilde{Q},B)_{\rho^2}=\left(-1,1\right),
		\end{aligned}
		\label{su2xu1_charges}
	\end{equation}
	for the projections of isospin and baryonic charge. 
	
	The global symmetry group of the \wcpt model is \eqref{glob_sym_N=2}, so
	the formula \eqref{Zmq} reads in this case,
	\begin{equation}
		\label{Zmq2}
		Z_{BPS} = i \vec{m}_n \vec{q}_n - i \vec{m}_\rho \vec{q}_\rho = i \frac{B}2\left(\sum m - \sum \tilde{m}\right) +  i \vec{m}_n \vec{q}_n^\bot - i \vec{m}_\rho \vec{q}_\rho^\bot
	\end{equation}
	Each \(\vec{q}_\bot\) corresponds to an \(\mathfrak{sl}_2\) charge via
	\begin{equation}
		\vec{q}_\bot = \sum_{i=1,2}q_i\vec{\nu}_i = \frac12 (q_1-q_2)\vec{\alpha} = \frac{Q}2 \vec{\alpha}= \mathsf{q} \vec{\alpha},
	\end{equation}
	where 
	\begin{equation}
		\vec{\nu}_1 = \vec{e}_1 - \frac12 \vec{e} = \frac12 (\vec{e}_1-\vec{e}_2) = \frac12 \vec{e} - \vec{e}_2 = - \vec{\nu}_2
	\end{equation}
	and \(\vec{\alpha} = \vec{e}_1-\vec{e}_2\). Formula \eqref{Zmq2} therefore gives
	\begin{equation}
		\label{Z2}
		Z_{BPS} = i \frac{B}2\left(m_1+ m_2 - \tilde{m}_1-\tilde{m}_2\right) +  i (m_1-m_2)  \mathsf{q} - i  (\tilde{m}_1-\tilde{m}_2)  \tilde{\mathsf{q}},
	\end{equation}
	for instance, $Z_{n^1\rho^2}=i(m_1-\tilde{m}_2)$ 	[cf. with \eqref{Z_pert}]. 
	
	Now let us turn to the nonperturbative states; the lightest is the massless stringy baryon, associated with the conifold complex structure parameter  $b$. 
	Eq. \eqref{deformedconi} shows that 
	the $b$-baryon  transforms with respect to \eqref{glob_sym_N=2} as \(\det w^{ij}\), namely, it is a singlet with respect to both SU(2) factors in \eqref{glob_sym_N=2} and has $B=2$.  
	This state is characterized by the $\mathfrak{u}_2$ charges  \eqref{u2_charges}, i.e.,
	\begin{equation}
		\vec{q}_{b}=(1,0)+(0,1)=(1,1)
	\end{equation} 
	or
	\begin{equation}
		Q=\tilde{Q}=(\vec{q}_{b})_1 -(\vec{q}_{b})_2=0,\ \ \ \ \ B=(\vec{q}_{b})_1+ (\vec{q}_{b})_2=2.
	\end{equation}
	With these $B$ and $Q$, Eq. \eqref{Z2} gives
	\begin{equation}
		Z_{b}=i(m_1+m_2-\tilde{m}_1-\tilde{m}_2).
		\label{Z_b}
	\end{equation}
	This result was obtained in \cite{ISY_b_baryon}. Note that within our interpolation procedure the mass of this state in the \wcpt model vanishes even at finite $\mathcal{M}$ because $\tilde{m}_i=m_i$; moreover, $m_1=m_2=0$.
	% 2D-4D correspondence tells us that
	%$b$-baryon in 4D SQCD is also massless.
	
	All massive string states in the theory with $N=2$ are also singlets with respect to both SU(2) factors in \eqref{glob_sym_N=2}, see Sec.~\ref{sec: 4D spectrum}.
	They differ only by the value of their baryonic charge $B$. Therefore, their central charges are just multiples of the one in \eqref{Z_b}. Thus, the BPS part of the mass is always zero for these states. They become massive due to the first non-BPS term in 
	\eqref{BPS+non-BPS}, which depends only on the string tension, $\tau=2\pi\xi$.

	\subsubsection{\(SU(4)\) case}
	
	Now consider our interpolation procedure at finite $\mathcal{M}$. In particular, at 
	$\mathcal{M}=0$ the global symmetry group reads
	\beq
	SU(4)_{n}\times SU(4)_{\rho}\times U(1)_B,
	\label{glob_sym_N=4}
	\eeq
	where each SU(4) factor is associated with $n^i$ and $\rho^j$ fields, respectively, $i,j=1,...,4$.
	At nonzero $\mathcal{M}$ each SU(4) factor is broken down to
	\beq
	SU(4)\to SU(2)_{1,2}\times SU(2)_{3,4}\times U(1),
	\label{SU(4)_breaking}
	\eeq
	see \eqref{mass_split}.
	The groups $SU(2)_{1,2}$ and $SU(2)_{3,4} $ rotate the fields $n^{1,2}$ ($\rho^{1,2}$) and   
	$n^{3,4}$ ($\rho^{3,4}$)  respectively, and therefore correspond to the \(a=1\) and \(a=3\) subgroups.

	To consider the nonperturbative string states, we have to decide first which quantum numbers with respect to the global group \eqref{glob_sym_N=4} are allowed. In the $N=2$ case all string states are singlets with respect to both SU(2) factors in \eqref{glob_sym_N=2}, but all have nonzero baryonic charges $B$. 
	Moreover, from the point of view of their transformation properties these states should behave as  powers of $\det w$ to be nontrivial on the deformed conifold, see \eqref{deformedconi}; this latter requirement (assuming, in particular, equality \(Q=\tilde{Q}\) of the \(n\)- and \(\rho\)-charges) we call a ``conifold rule''.
	These arguments suggest that we  propose the following generalization of these rules~\footnote{Below in this section we present only some qualitative features of the hadron states in \(SU(4)\) gauge theory. We postpone a more detailed analysis of this picture in a generic \(SU(N)\) case for a separate publication.} for the \wcpf model: 
	
	(i) First, we assume that 
	all allowed  stringy states at nonzero $\mathcal{M}$ are singlets with respect to both
	$SU(2)_{1,2}$ and $SU(2)_{3,4}$ subgroups in \eqref{SU(4)_breaking} for both SU(4) groups associated with $n$ and $\rho$ fields, see \eqref{glob_sym_N=4}.   It means that
	\begin{equation}
		\label{singlet13}
		Q_1=Q_3=\tilde{Q}_1=\tilde{Q}_3=0.
	\end{equation}
	In contrast to the \(N=2\) case, the $b$-baryon
	now belongs to a nontrivial (long fundamental) representation {\bf 6} in $\bf{4}\times \bf{4} = \bf{10} +\bf{6}$ of SU(4) with the highest weight \(\vec{\mu}_2=\vec{\nu}_1+\vec{\nu}_2\).
	Hence, a stronger requirement for all states to  be singlets of the whole SU(4) would kill the $b$-baryon state, which is expected, however, to survive the transition to a theory with a larger gauge group. 
	
	(ii) Second, we still assume a generalized ``conifold rule'',  requiring that the charges of a stringy state with respect to the $SU(4)_{\rho}$ and $SU(4)_{n}$ groups coincide,
	i.e., in addition to \eqref{singlet13}, impose
	\begin{equation}
		Q_2=\tilde{Q}_2=Q
	\end{equation}
	being the charge of the diagonal \(SU(2)_{2,3}\subset SU(4)_d\subset SU(4)_{n}\times SU(4)_{\rho}\)

	Formula \eqref{ZmqC} then gives for the baryons,
	\begin{equation}
		\begin{aligned}
			Z_{BPS} &=
			i \frac{B}4\left(\sum_{j=1}^4 m_j - \sum_{j=1}^4 \tilde{m}_j\right) +  i \sum_{a,b=1}^{3}\left( (m_a-m_{a+1}-\tilde{m}_a+\tilde{m}_{a+1})C^{-1}_{a2}Q
			\right) =
			\\
			&= i(m_1+m_2-\tilde{m}_1-\tilde{m}_2)\left(\frac{B}4+\frac{Q}2\right)  + i(m_3+m_4-\tilde{m}_3-\tilde{m}_4)\left(\frac{B}4-\frac{Q}2\right) 
		\end{aligned}
		\label{Zmq4}
	\end{equation}
	Consider first the $b_{1,2}$-baryon, which is a particular component $(w^{11}w^{22}-w^{12}w^{21})$
	of ${\bf 6}_{n}\times {\bf 6}_{\rho}$ with $u(4)$ charges
	\begin{equation}
		\vec{q}_{b_{1,2}}=\vec{q}_{1}+\vec{q}_{2}=(1,0,0,0)+(0,1,0,0)=(1,1,0,0)
	\end{equation} 
	For $\{Q_a\}$ and the baryonic charge one gets
	\begin{equation}
		\begin{aligned}
			&Q_1=(\vec{q}_{b_{1,2}})_1-(\vec{q}_{b_{1,2}})_2=1-1=0,\\
			&Q_3=(\vec{q}_{b_{1,2}})_3-(\vec{q}_{b_{1,2}})_4=0-0=0,\\
			&Q_2\equiv Q=(\vec{q}_{b_{1,2}})_2-(\vec{q}_{b_{1,2}})_3=1-0=1\\
			& B=\sum_j(\vec{q}_{b_{1,2}})_j=1+1+0+0=2.
		\end{aligned}
	\end{equation}
	Similarly, for the $b_{3,4}$ state, which is another component $(w^{33}w^{44}-w^{34}w^{43})$ of ${\bf 6}_{n}\times {\bf 6}_{\rho}$, we have 
	\begin{equation}
		\begin{aligned}
			&Q_1=(\vec{q}_{b_{3,4}})_1-(\vec{q}_{b_{3,4}})_2=0-0=0,\\
			&Q_3=(\vec{q}_{b_{3,4}})_3-(\vec{q}_{b_{3,4}})_4=1-1=0,\\
			&Q_2\equiv Q=(\vec{q}_{b_{3,4}})_2-(\vec{q}_{b_{3,4}})_3=0-1=-1,\\
			&B=\sum_j(\vec{q}_{b_{3,4}})_j=0+0+1+1=2.
		\end{aligned}
	\end{equation}
	Hence, one gets here two baryons ($b_{1,2}$ and  $b_{3,4}$) with charges \(Q=\pm 1\) with respect to the    \(\vec{\alpha}_2\)-subgroup \(SU(2)_{2,3}\subset SU(4)\), respectively. Their masses are
	\beq
	Z_{b_{1,2}}=i \left(m_1+m_2- \tilde{m}_1-  \tilde{m}_2\right).
	\label{Z_b_12}
	\eeq
	and
	\beq
	Z_{b_{3,4}}=i \left(m_3+m_4- \tilde{m}_3-  \tilde{m}_4\right)
	\label{Z_b_34}
	\eeq
	as expected. 
	
	Geometrically, all states with \(\mathfrak{u}_4=(\mathfrak{su}_4)_d\oplus (\mathfrak{u}_1)_B\) charges form an integer lattice \(\mathbb{Z}^4\subset \mathbb{R}^4\). The \((\mathfrak{su}_2)_{1,2}\times (\mathfrak{su}_2)_{3,4}\) singlet states we are looking for  form a sublattice \(\mathbb{Z}^2\subset \mathbb{R}^2\), generated by the (orthogonal) vectors \((\vec{e}_1+\vec{e}_2,\vec{e}_3+\vec{e}_4)\), i.e.
	\beq
	\vec{q} = l\,\vec{q}_{b_{1,2}}+ k\, \vec{q}_{b_{3,4}} = l(\vec{e}_1+\vec{e}_2) + k(\vec{e}_3+\vec{e}_4),
	\eeq
	where $l$ and $k$ are non-negative integers, see Fig.~\ref{fig:new_states}. The baryonic charge of these states is measured by
	\begin{equation}
		B =  \vec{e}\cdot \vec{q} = 2 (k+l)
	\end{equation}
	and the BPS masses of these stringy states are given by
	
	\begin{equation}
		\begin{aligned}
			&Z_{BPS} = i (\vec{m}_n - \vec{m}_\rho )\vec{q} \\
			&=  i \left[ l\,\left(m_1+m_2- \tilde{m}_1-  \tilde{m}_2 \right)+k\,\left(m_3+m_4- \tilde{m}_3-  \tilde{m}_4\right)\right],\quad
			l+k=\frac{B}{2}.
		\end{aligned}
		\label{Z_BPS}
	\end{equation}

	\subsection{Number of states}

	\begin{figure}[t]
		\centering
		\includegraphics[width=0.9\linewidth]{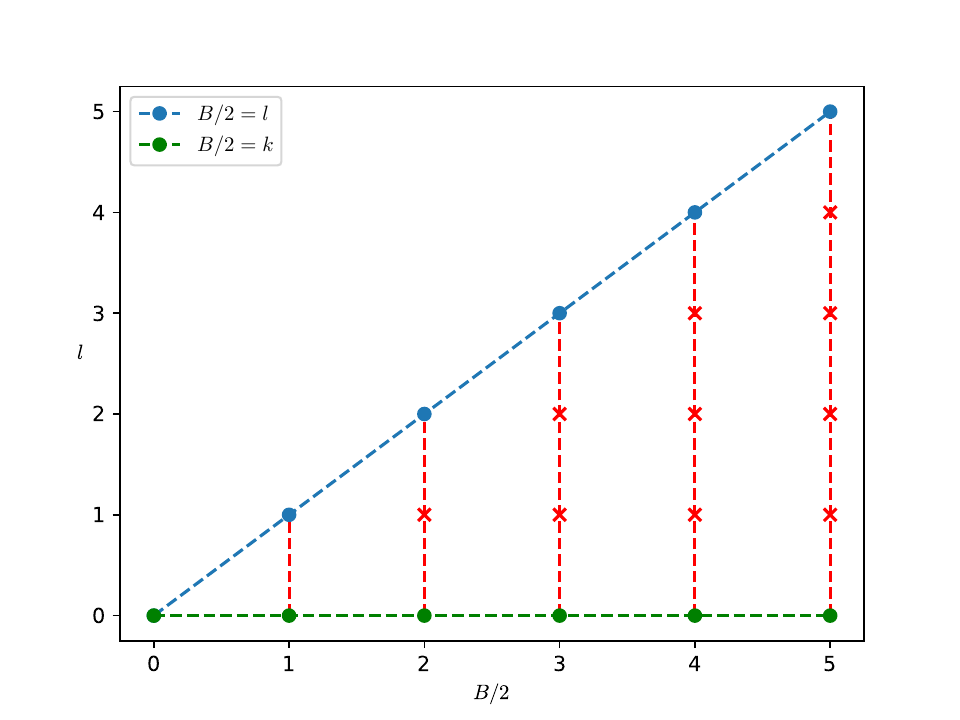}
		\caption{
			States with different baryonic charges. The blue dashed line corresponds to $l=\frac{B}{2},k=0,$ the green one to $l=0,k=\frac{B}{2}$. Red crosses are new states that emerge as we reduce $\mathcal{M}$.
		}
		\label{fig:new_states}
	\end{figure}
	
	Actually, the main observation we need from the previous Sec.~\ref{sec:central_charges} is that under 
	our mass deformation we always have $\tilde{m}_i=m_i$, and therefore the BPS parts of the masses
	of all these states in \eqref{BPS+non-BPS} vanish for the states satisfying the ``conifold rule''. This is a simple consequence of the general formula \eqref{Zmq}, and one can easily see this effect in the considered above examples \eqref{Z_b_12}, \eqref{Z_b_34} and for the generic case \eqref{Z_BPS}.
	
	It actually means that the form of the spectrum \eqref{BPS+non-BPS} does \emph{not} change as we reduce the mass deformation parameter $\mathcal{M}$. 
	However, this is not so trivial, since the \emph{number} of physical states changes for each fixed baryon charge $B$. Namely, at $\mathcal{M}\to\infty$, with the infinite mass barrier, certain quarks totally decouple from the theory. In fact, in this limit the theory factorizes into
	two noninteracting SQCDs as shown in \eqref{M_to_infty} for $K=2$,
	and there is no way to form a state with mixed $b_{1,2}$ and $b_{3,4}$  global charges. 
	Hence, at $\mathcal{M}\to\infty$ we have only two states with central charge given by \eqref{Z_BPS}, corresponding to
	$l=B/2$, $k=0$  and/or $l=0$, $k=B/2$,
	% i.e. with the charges of massless quarks, 
	while all other states are separated by infinite barriers. 
	Instead, for finite $\mathcal{M}$ at each 
	value of $B$ more states are allowed, actually, at small $\mathcal{M}$, all of them with $l+k=B/2$, so that the whole ``cone'' in Fig.~\ref{fig:new_states} with internal points, marked with red, appears in the spectrum, increasing the number of states with the same mass.
	
	We also expect that, as we reduce $\mathcal{M}$, the system goes through a certain number of crossovers, similar to walls of marginal stability for BPS states, where the number of states jumps. In the next section we will see that
	this behavior is in qualitative agreement with results from string theory.

	\section{Number of states from black hole}
	\label{sec:Hagedorn}
	\setcounter{equation}{0}

	\subsection{Hagedorn behavior}
	Now let us return to  the string theory described by \ntwo Liouville world sheet model \eqref{deformed_Liouville} with both Liouville superpotential and mass deformation switched on.  As we discussed in Sec. \ref{sec:superpot+mass}, this theory is dual to the 2D black hole  \eqref{BH} with a cigar geometry.   We also conjectured that the mass of the black hole is given by Eq. \eqref{M_BH_total} and depends on both deformation parameters $b$ and $\mathcal{M}$. We will show in this section that, although
	the string spectrum does not change as we reduce the mass deformation parameter 
	$\mathcal{M}$ (which has a reasonable explanation on the field theory side, see the previous section), the number of states on each energy level increases.
	
	One of the ways to find the number of  states is to calculate the entropy of the system. 
	In Euclidean formulation, the compact dimension of the target space can be interpreted as a temperature $T=(2\pi R)^{-1}$, where $R=\sqrt{2k}$ is the asymptotic radius of the cigar, see \eqref{BH}. We have to stress here that in our theory real time is one of the 4D Minkowski coordinates and has nothing to do with the compact dimension around the cigar, but we use the  trick with the Euclidean 2D black hole to estimate the  number of states in 4D theory.
	
	This idea, however, immediately faces a problem. In string theory there is a limiting Hagedorn   temperature $T_H$ \cite{Hagedorn}  beyond which the higher  energy levels are no longer suppressed by the factor $\exp{(-E/T)}$ due to the exponential growth of the density of the string states  
	\beq
	\omega(E) \sim \exp{\left(\frac{E}{T_H}\right)}
	\label{density_states}
	\eeq
	and the partition function 
	\beq
	Z = \int_{0}^{\infty} dE \,\omega(E)\, e^{-\frac{E}{T}}
	\eeq
	becomes divergent  at $T \geq T_H$ (see \cite{AtickWitt} and references therein). Similarly, the exponential growth \eqref{density_states} leads to another unusual fact \cite{FM}, that string theory belongs to a nonlocalizable class of theories where the Green functions can have singularities in a finite space-time domain (of the order of the string length).
	
	In the Euclidean formulation, the Hagedorn behavior is described by the so-called thermal scalar, which is a winding string mode around the thermal circle \cite{AtickWitt}. As temperature approaches $T_H$ this mode becomes massless and presumably becomes tachyonic at $T>T_H$, leading to an instability.
	
	For the black holes we are interested in here, the Hagedorn behavior is related to the black hole/excited strings transition  \cite{Susskind,HorowPolch}.  At low temperatures, we have a well-defined black hole geometry with small $\alpha'$ corrections. As the temperature grows above some critical value, the string's size exceeds its Schwarzschild radius and the black hole turns into an excited string \cite{HorowPolch}.
	
	Similar behavior was found for the 2D black hole  with the linear dilaton \eqref{BH}  in \cite{GivKutRabin}. As we reduce $k$ (increase temperature),  the theory enters a strong coupling regime where  $\alpha' \sim 1/k$ corrections grow, and below some critical value $k_c$,  the black hole, as a geometric object, no longer exists. Moreover, it is argued in \cite{GivKutRabin} that the critical value $k_c=1$ is equal to unity, i.e., exactly the value we are interested in for the critical non-Abelian string in 4D SQCD. At this value, the thermal winding scalar becomes  massless \cite{GivKutRabin,Kut_wind_condens}, see also \cite{KazakKostovKut} for the nonsupersymmetric version. Below, in this section, we consider $k$ above the critical value $k = 1$ to regularize the theory. We use the target-space effective action to calculate the entropy, associated with the thermal scalar, and find its dependence on the mass deformation parameter $\mathcal{M}$.
	
	\subsection{Thermal scalar}
	
	The thermal scalar for the 2D black hole \eqref{BH} is our (massless in 4D) $b$-baryon, associated with the conifold complex structure parameter $b$.  It corresponds to the minimal winding $n_2=\pm1$ state, see \eqref{m} \cite{AtickWitt}, i.e., its quantum numbers are
	\beq
	m=\pm \frac{k}{2}, \qquad j=-\frac{k}{2}. 
	\label{j_m_b}
	\eeq
	This state belongs to the discrete spectrum, see Sec.~\ref{sec:vertices}; for $k>1$ this state is normalizable and for $k=1$ it is on the borderline between normalizable and non-normalizable modes. It is  seen from the asymptotic behavior~\footnote{This behavior was found in \cite{Kut_wind_condens}, see also \cite{KazakKostovKut,Malda_BH_hagedorn} for the nonsupersymmetric version.}
	\beq
	\Psi^{(t)}_b(\rho)\ \stackreb{\rho\to\infty}{\approx}  e^{2\left(j+\frac12\right)\rho}=  e^{-(k-1)\rho},
	\label{Psi_b_infty}
	\eeq
	of the wave functions \(\Psi^{(t)}_b(\rho)= \Psi^{(t)}_{-k/2,\pm k/2}(\rho)\), see \eqref{Psi_t}.  It shows that at $k\to 1$ the thermal scalar  becomes massless (with respect to  the $\rho $ dimension), but never becomes tachyonic, $-E_j^{(t)} \ge 0$,  see  \eqref{E_t}.   
	
	Its effective theory can be described by the tachyonic action \eqref{tachyon_action} in the supergravity background \eqref{gravity_action}. For the tachyon \eqref{dressed_tachyon} with quantum numbers \eqref{j_m_b} on mass-shell, when the conformal dimension \eqref{dimV} satisfies
	\beq
	2\Delta_{j,m}(T_b)= 2\Delta_{-k/2,\pm k/2}=1,
	\label{b_mass_shell}
	\eeq
	the action \eqref{tachyon_action} can be reduced to 2D Euclidean theory,
	\beq
	S_{b}^E= \frac1{2\tilde{\kappa}^2}\int d \rho d \vartheta\, \sqrt{G}\,e^{-2\Phi}\,\left\{g^{nn'}\pt_{n}\bar{T}_{b}\,\pt_{n'} T_{b} 
	- |T_{b}|^2\right\},
	\label{b_action_E}
	\eeq
	where $\frac1{2\tilde{\kappa}^2}=\frac{V_{4D}}{2{\kappa}^2}=\frac{V_{4D}}{(2\pi)^3\alpha'^{2}}$ is proportional to the volume of the 4D Minkowski space, while the string coupling $g_s^2=\exp{(2\Phi_0)}$ is included into $\exp{(-2\Phi)}$.  
	
	Since our thermal scalar ($b$-baryon) is a winding mode in cigar theory, \eqref{b_action_E} is written in the $T$-dual trumpet metric,
	\eqref{trumpet}
	\beq
	ds^2=2k\left(d\rho^2 +\coth^2{\rho}\, d\vartheta^2\right),
	\label{trumpet_metric}
	\eeq
	with the dilaton \eqref{dilaton_rho}, see \cite{Giveon,DijVerVer}.
	
	To calculate the entropy of the black hole we use the formula for the (dominant) part of the black hole entropy, associated with the thermal scalar
	\beq
	s=\left( 1- R\frac{\pt}{\pt R}\right) \log{Z} \approx \left(  R\frac{\pt}{\pt R} -1 \right) S_b^E,
	\label{entropy_gen}
	\eeq
	in the quasiclassical limit (see 
	review \cite{Witt_BH_rev} for a recent discussion). Here $R=\sqrt{2k}$ is the radius of the cigar at $\rho\to\infty$, proportional to the inverse Hawking temperature.
	However, one should use the formula \eqref{entropy_gen} with great care, since changing the level \(k\) not only deforms the radius of the ``temperature circle'' but also many properties of the whole theory, in particular it rescales the radial direction in the background \eqref{trumpet_metric}, see \cite{Malda_BH_hagedorn} for the discussion of this peculiarity for the case of the nonsupersymmetric 2D black hole.
	
	In order to avoid this problem, we first take the action \eqref{b_action_E} after the substitution 
	\beq
	T_b (\rho, \theta) = e^{2im \vartheta} \mathsf{T}(\rho)\ \stackreb{m=\pm R^2/4}{=}\ e^{\pm i\frac{R^2}{2} \vartheta} \mathsf{T}(\rho)
	\label{tildeT},
	\eeq
	together with ($k$-dependent!) metric \eqref{trumpet_metric} and dilaton \eqref{dilaton_rho}. After
	the integration over the compact direction \(\vartheta\sim\vartheta+\frac{4\pi}{R^2}\) [we separate the radius and the level $k$ for a moment, cf. with \eqref{Y_t_period}]  one gets 
	\beq
	S_{b}^E= \frac{2\pi e^{-2\Phi_0}k}{\tilde{\kappa}^2R^2}\int_0^\infty d \rho \sinh(2\rho)\left\{\frac1{2k}(\d_\rho \mathsf{T})^2+\left(\frac{R^4}{8k}\tanh^2\rho-1\right)\mathsf{T}^2\right\}.
	\label{b_action_rho}
	\eeq
	To find the entropy one substitutes \eqref{b_action_rho} into \eqref{entropy_gen}, where the derivative is taken only of the explicit \(R\)-dependence, since the possible extra terms, coming from the derivatives of the fields, vanish, being proportional to the equations of motion~\footnote{For the thermal scalar  \(\mathsf{T}\) equation of motion is equivalent to \eqref{Psi_eq_t} at \(j=-k/2\), \(m=\pm k/2\) under \(\mathsf{T}(\rho)=\frac{\Psi_{-k/2,\pm k/2}(\rho)}{\sqrt{\sinh 2\rho}}\), while in the equations of motion for the background \eqref{trumpet_metric} and \eqref{dilaton_rho} we neglect the contribution of the field \(\mathsf{T}\).}, and we keep \(k\) constant when taking \(R\)-derivatives, putting \(R=\sqrt{2k}\) only at the end.  This leads to
	\begin{equation}
		\begin{aligned}
			&s= \left.\left(  R\frac{\pt}{\pt R} -1 \right) S_b^E\right|_{R=\sqrt{2k}}=\\
			&=
			\frac{\pi e^{-2\Phi_0}}{\tilde{\kappa}^2}\int_0^\infty d \rho \sinh(2\rho)\left\{-\frac3{2k}(\d_\rho \mathsf{T})^2+\left(\frac{k}2\tanh^2\rho+3\right)\mathsf{T}^2\right\},
		\end{aligned}
		\label{b_entropy_int}
	\end{equation}
	Substituting here  \(\mathsf{T}(\rho)=\frac{\Psi_{-k/2,\pm k/2}(\rho)}{\sqrt{\sinh 2\rho}} = \frac{1}{\sqrt{2}\cosh^k\rho}\) and computing the integral, one finds
	\beq
	\frac{(2\pi \alpha')^2}{V_{4D}}\,s = \frac{e^{-2\Phi_0}}{k-1} = \frac{1}{k-1} \left(e^{-2\Phi_0^{(b)}} + e^{-2\Phi_0^{(\mathcal{M})}}\right)
	=  \frac{\sqrt{2k}}{k-1}M_{BH}^{\rm total}
	\label{b_entropy}
	\eeq
	where in the last two equalities we have used \eqref{phi_total} and \eqref{M_BH_total}. Hence, we see that the entropy \eqref{b_entropy} diverges at $k\to 1$, a similar result for the entropy of the nonsupersymmetric 2D black hole was obtained in \cite{Malda_BH_hagedorn}. This divergence comes from 
	the infrared region $\rho\to\infty$, due to asymptotic behavior  \eqref{Psi_b_infty}, leading to explosion of the norm \(\int d\rho \sinh(2\rho)|\mathsf{T}|^2=\int d\rho|\Psi_b|^2\) of the thermal scalar wave function at \(k\to 1\).
	
	The rest of the contributions into the black hole entropy (not associated with the thermal scalar) are also proportional to $M_{BH}^{\text{total}}=\frac{S_{BH}^{\text{total}}}{2\pi\sqrt{2k}}$, but give subleading terms
	in the limit \(k\to 1\). As argued in \cite{GivKutRabin,Kut_wind_condens} the critical value $k_H=1$ is
	related to the black hole/excited strings phase transition.
	
	Note that the Hagedorn temperature in flat space in the supersymmetric case corresponds to $k_H^{\rm flat}=2$, see \cite{Mertens} for a review.  Still, in the cigar geometry one can increase the temperature till $k$  reaches the critical value $k_H=1$ from above, similar behavior for the 
	nonsupersymmetric case was discussed in \cite{Malda_BH_hagedorn}
	
	Observe now that due to \eqref{M_BH_total} at the initial stage of the mass deformation at 
	$\mathcal{M}\to\infty$ the entropy \eqref{b_entropy} does not really feel
	the mass parameter, being just $s\sim |b|^2(k-1)^{-1}$. However, at small $\mathcal{M}\to 0$ the black hole mass increases due to the second term in \eqref{M_BH_total}, giving a ``second divergence'' to the entropy, becoming
	\beq
	\frac{(2\pi \alpha')^2}{V_{4D}}\,s \sim \frac{1}{k-1}\,\frac{|b|^2}{\mathcal{M}^2}.
	\label{b_entropy_mu}
	\eeq
	Hence, we always meet the Hagedorn behavior of excited strings, which implies that all high energy levels are filled now in the system. However, we also see that the number of states  (finite at each  level with fixed energy or baryonic charge $B$) increases dramatically as we reduce $\mathcal{M}$. 
	This qualitatively confirms our expectations on the field theory side. Our  4D \ntwo supersymmetric  theory interpolates  from SQCD with the gauge group $U(2)$ and $N_f=4$ to SQCD with the gauge group $U(4)$ and $N_f=8$ as we reduce the mass deformation parameter $\mathcal{M}$, and one naturally expects more hadronic states to appear in a theory with more quark matter multiplets.

	\section{Conclusions}
	\label{sec:Conclusions}
	\setcounter{equation}{0}
	
	In this work we examined a particular mass deformation of $\mathcal{N}=2$ SQCD with gauge group U(4) and $N_f=8$ hypermultiplets of fundamental quarks,
	introducing a bare mass $\mathcal{M}$ for half of the quark flavors, while keeping the other half without a mass term.
	At large $\mathcal{M}$, this theory splits into two non-interacting sectors, each representing an SQCD with two colors and four flavors.
	As reviewed in Sec.~\ref{sec: 4D spectrum}, the hadron spectrum of the latter theory is known from the string theory of the critical non-Abelian string \cite{SYlittles,SYlittmult}.
	% due to 2D-4D correspondence with solvable world-sheet theory.
	
	Perhaps our most surprising result is that the spectrum of string states of the mass-deformed theory does \textit{not} change  when we reduce the mass parameter 
	$\mathcal{M}$ interpolating  to SQCD with  more quark flavors. In particular, the low-lying spectrum of hadrons in 4D SQCD is still given by \eqref{tachyonmass} and \eqref{gravitonmass} and is actually independent of $\mathcal{M}$. 
	
	On the string theory side, this fact is established with the help of the deformed \ntwo Liouville theory \eqref{deformed_Liouville} on the world sheet of the non-Abelian strings supported in \ntwo SQCD.
	First, we  proved that the mass-deformed theory with the Liouville superpotential dropped in \eqref{deformed_Liouville} is in fact $T$-dual to the  \ntwo black hole with 
	the same cigar geometry.
	Then, combining mirror and $T$-duality arguments, we concluded that the theory \eqref{deformed_Liouville}
	with both deformations switched on is actually dual to the 2D black hole with increasing mass, determined by the deformation parameters.
	We also used field theory arguments on the  SQCD side to explain this surprising 
	behavior of the string spectrum.
	
	Nevertheless,  as a result of the deformation, the multiplicities of the hadron SQCD states with given baryonic charges increase as we reduce $\mathcal{M}$ and interpolate to SQCD with more quarks. We estimated the entropy of the resulting black hole near the Hagedorn transition, showing that the string theory results qualitatively confirm our field theory expectations.

	We have also derived the string spectrum, associated with the mass-deformed  Liouville theory,  explicitly solving the equations for vertex operators' wave functions in the gravity background. Our result for the reflection coefficient (which fixes the discrete spectrum)
	coincides with the known exact result for the 2D black hole in the large $k$ limit. This matches
	our conclusion that the two theories are just related by $T$-duality. It also shows that the effective gravity approach for the reflection coefficient misses the  $\alpha'\sim 1/k$ correction even in the supersymmetric case, where it reproduces the metric and dilaton exactly.

	\section*{Acknowledgments}
	
	The authors are grateful to  A.~Litvinov and A.~Sidorenko  for useful and 
	stimulating discussions. The work of G.S. and A.Y.   was  supported  by the Foundation for the Advancement of Theoretical Physics and Mathematics ''BASIS'',  Grant No. 22-1-1-16. The work of E.I. was supported in part by U.S. Department of Energy Grant No. de-sc0011842. The work of A.M. was supported by the Basic Research Program of HSE University, A.M. is also grateful to BIMSA, where an essential contribution to this work has been done. The work of G. S. was also partly supported by the Gribov Scholarship for works in the field of theoretical physics.

	\appendix

	\section{The analytic continuation of the singular solution at $z=0$ to $z=-\infty$}
	\label{app:sing_cont}

	The second, singular at $z=0$ solution to Eq. \eqref{w_eq_z} takes the form
	%\begin{equation}
	%		w_{sing}=F(\alpha,\beta,1;1-y)\ln(1-y) + \text{polynomial in $(1-y)$}.
	%\end{equation}
	%
	%The second linear independent  solution, singular at $z=0$ has the form
	\begin{equation}
		\begin{aligned}
			w_{sing}
			&= F(\alpha,\beta,1,z)\log z \\ 
			&+ \sum_{k=1}^{\infty}(z)^k\frac{(\alpha)_k(\beta)_k}{(k!)^2}\{\psi(\alpha+k)-\psi(\alpha)+\psi(\beta+k)-\psi(\beta)-2\psi(k+1)+2\psi(1)\}, 
		\end{aligned}
	\end{equation}
	where $(a)_n=a(a+1)...(a+n-1)$ is the Pochhammer symbol.
	
	To make an analytic continuation of this solution to infinity, consider Eq. (9.131.2) from \cite{Ryzhik} and apply it to our case $\gamma-\alpha-\beta=\epsilon$ with $\epsilon\rightarrow 0$,
	\begin{equation}
		\begin{aligned}
			F(\alpha,\beta,\gamma;1-z)&=\frac{\Gamma(\gamma)\Gamma(\gamma-\alpha-\beta)}{\Gamma(\gamma-\alpha)\Gamma(\gamma-\beta)}F(\alpha,\beta,\alpha+\beta-\gamma+1;z)\\
			&+z^{\gamma-\alpha-\beta}\frac{\Gamma(\gamma)\Gamma(\alpha+\beta-\gamma)}{\Gamma(\gamma-\alpha)\Gamma(\gamma-\beta)}F(\gamma-\alpha,\gamma-\beta,\gamma-\alpha-\beta+1;z)\\
			&=\lim\limits_{\epsilon\rightarrow 0}\frac{\Gamma(\gamma)}{\Gamma(\beta)\Gamma(\alpha)}\left(\Gamma(\epsilon)F(\alpha,\beta,1-\epsilon;z)+z^{\epsilon}\Gamma(-\epsilon)F(\beta,\alpha,1+\epsilon;z)\right)\\
			&=\frac{\Gamma(\gamma)}{\Gamma(\beta)\Gamma(\alpha)}F(\alpha,\beta,1;z)\lim\limits_{\epsilon\rightarrow 0}(\Gamma(\epsilon)+z^\epsilon\Gamma(-\epsilon))\\
			&=-\frac{\Gamma(\gamma)}{\Gamma(\beta)\Gamma(\alpha)}F(\alpha,\beta,1;z)(2\gamma_e+\ln z),
		\end{aligned}
		\label{sing_prod}
	\end{equation}
	where  $\alpha$ and $\beta$  are given by \eqref{alpha_beta}, $\gamma=1-2m$, while $\gamma_e$ is the Euler's constant. In \eqref{sing_prod}, when moving to line four we use the symmetry of the hypergeometric function under interchange of its first two parameters ($\alpha$ and  $\beta$), and in the last line we use the following expansion:
	\begin{equation}
		\Gamma(\epsilon)+(z)^\epsilon\Gamma(-\epsilon)\stackrel{\epsilon\rightarrow 0}{\approx}-(2\gamma_e+\ln z)+O(\epsilon).
		\label{eps_expansion}
	\end{equation}
	Hence, neglecting the $\sim \gamma_e$ term, from \eqref{sing_prod} we get
	\begin{equation}
		F(\alpha,\beta,1;z)\ln z\approx-\frac{\Gamma(\beta)\Gamma(\alpha)}{\Gamma(\gamma)}F(\alpha,\beta,\gamma;1-z).
		\label{from1to0}
	\end{equation}
	This means that the singular solution at $z=0$ behaves as the regular one at $z=1$. Using this property, we make an analytic continuation to infinity,
	\begin{equation}
		\begin{aligned}
			w_{sing}&\approx-\frac{\Gamma(-j-m)\Gamma(1+2j)}{\Gamma(1+j-m)}(z-1)^m
			\times\bigg[(z-1)^j F\left(-j-m,-j+m,-2j;\frac{1}{1-z}\right) \\
			&+R_{sing} (z-1)^{-1-j} F\left(1+j-m,1+j+m,2+2j;\frac{1}{1-z}\right)  \bigg],
		\end{aligned}
	\end{equation}
	where

	\begin{equation}
		R_{sing}=\frac{\Gamma(-1-2j)\Gamma^2(1+j-m)}{\Gamma(1+2j)\Gamma^2(-j-m)}.
		\label{R_sing_noR}
	\end{equation}
	Obviously, $R_{sing}$ has no symmetry under the replacement $m\rightarrow -m$, which contradicts our condition $m\equiv m_L=-m_R$, see footnote~\ref{foot:mLR} on page~\pageref{foot:mLR}.

	%%%%%%%%%%%%%%%%%%%%%%%%%%%%%%%%%%%%%%%%%%%%%%%%%%%%%%%%%%%%%%%%%%%%%%%%%%%%%%%%%%%%%%%%%%%%%%%%%%%%%%%%%%%%%%%%%%%%%%%%%%%%%%%%%%%%%%%%%%%%%%%
	%%%%%%%%%%%%%%%%%%%%%%%%%%%%%%%%%%%%%%%%%%%%%%%%%%%%%%%%%%%%%%%%%%%%%%%%%%%%%%%%%%%%%%%%%%%%%%%%%%%%%%%%%%%%%%%%%%%%%%%%%%%%%%%%%%%%%%%%%%%%%%%
	%
	%                            B I B L I O G R A P H Y
	%
	%%%%%%%%%%%%%%%%%%%%%%%%%%%%%%%%%%%%%%%%%%%%%%%%%%%%%%%%%%%%%%%%%%%%%%%%%%%%%%%%%%%%%%%%%%%%%%%%%%%%%%%%%%%%%%%%%%%%%%%%%%%%%%%%%%%%%%%%%%%%%%%
	%%%%%%%%%%%%%%%%%%%%%%%%%%%%%%%%%%%%%%%%%%%%%%%%%%%%%%%%%%%%%%%%%%%%%%%%%%%%%%%%%%%%%%%%%%%%%%%%%%%%%%%%%%%%%%%%%%%%%%%%%%%%%%%%%%%%%%%%%%%%%%%
	
	%\vspace{1cm}
	
	%\clearpage
	
	%\section*{TEMPORARY LABEL [REMOVE THIS IN THE FINAL VERSION]}.
	
	\renewcommand{\theequation}{A.\arabic{equation}}
	\setcounter{equation}{0}

\end{document}